\documentclass[reprint,NumberedRefs]{JASA}
\usepackage{xcolor}
\usepackage{multirow}
\newcommand*\patchAmsMathEnvironmentForLineno[1]{%
  \expandafter\let\csname old#1\expandafter\endcsname\csname #1\endcsname
  \expandafter\let\csname oldend#1\expandafter\endcsname\csname end#1\endcsname
  \renewenvironment{#1}%
     {\linenomath\csname old#1\endcsname}%
     {\csname oldend#1\endcsname\endlinenomath}}%
\newcommand*\patchBothAmsMathEnvironmentsForLineno[1]{%
  \patchAmsMathEnvironmentForLineno{#1}%
  \patchAmsMathEnvironmentForLineno{#1*}}%
\AtBeginDocument{%
\patchBothAmsMathEnvironmentsForLineno{equation}%
\patchBothAmsMathEnvironmentsForLineno{align}%
\patchBothAmsMathEnvironmentsForLineno{flalign}%
\patchBothAmsMathEnvironmentsForLineno{alignat}%
\patchBothAmsMathEnvironmentsForLineno{gather}%
\patchBothAmsMathEnvironmentsForLineno{multline}%
}

\begin{document}
\title[An explicit granular-mechanics approach to sediment acoustics]{An explicit granular-mechanics approach to marine sediment acoustics}
\author{Abram H. Clark}
\email{abe.clark@nps.edu}
\affiliation{Physics Department, Naval Postgraduate School, Monterey, CA 99343}
\author{Derek R. Olson}
\email{derek.olson@nps.edu}
\affiliation{Oceanography Department, Naval Postgraduate School, Monterey, CA 99343}
\author{Andrew J. Swartz}
\affiliation{Physics Department, Naval Postgraduate School, Monterey, CA 99343}
\author{W. Mason Starnes}
\affiliation{Physics Department, Naval Postgraduate School, Monterey, CA 99343}
\date{\today}

\begin{abstract}
Here we theoretically and computationally study the frequency dependence of phase speed and attenuation for marine sediments from the perspective of granular mechanics. We leverage recent theoretical insights from the granular physics community as well as discrete-element method simulations, where the granular material is treated as a packing of discrete objects that interact via pairwise forces. These pairwise forces include both repulsive contact forces as well as dissipative terms which may include losses from the fluid as well as losses from inelasticity at grain-grain contacts. We show that the structure of disordered granular packings leads to anomalous scaling laws for frequency-dependent phase speed and attenuation that do not follow from a continuum treatment. Our results demonstrate that granular packing structure, which is not explicitly considered in existing models, may play a crucial role in a complete theory of sediment acoustics. While this simple approach does not explicitly treat sound propagation or inertial effects in the interstitial fluid, it provides a starting point for future models that include these and other more complex features.
\end{abstract}

\maketitle
\section{Introduction}
\label{sec:intro}
The dispersion relation for acoustic waves is a fundamental observable for materials. The functions $c(f)$ and $\alpha(f)$ connecting the phase speed $c$ and attenuation coefficient $\alpha$ to the frequency $f$ depend strongly on the microscopic processes that control wave propagation. Marine sediments are fundamentally fluid-saturated granular materials, and there are many complex physical aspects to consider, even for an idealized scenario (e.g., neglecting biological matter~\cite{ballard2017acoustics}). For example, the nonlinear Hertzian~\cite{daraio2005strongly,gomez2012shocks} and inelastic contact forces~\cite{force_schemes} between the grains, the structure and other properties of the granular contact network~\cite{majmudar2005contact,Silbert_PRL_2005}, and the details of the interactions with the interstitial fluid~\cite{chotiros2017Acoustics,Williams2009b} all may affect the acoustic response. 

One of the important applications of dispersion models is their use in acoustic remote sensing of the seafloor, which commonly uses measurements of the reflection coefficient \cite{Holland2012,Dettmer2010,Quijano2012,Holland2017,Olson2023} or waveguide dispersion \cite{Bonnel2013,tan_ambient_2022}, combined with an acoustic model, to estimate the sound speed and density as a function of depth within the seafloor. If these measurements are made over a wide frequency band, then dispersion will be important \cite{Holland2013}. Another application of sediment dispersion models is the prediction of the bottom-interacting acoustic field due to a source  (i.e. propagation\cite{Porter1985} or reverberation\cite{Holland1998,LePage2000,LePage2003,Tang2017,Olson2020,Jackson2020,Olson2020a}), with the goal of performance modeling of remote sensing of objects in the ocean \cite{Abraham2019}. For these applications, linking the microscopic properties with the observables is a key step.

Real-world data for $c$ and $\alpha$ can be difficult to obtain for a wide range of $f$ and there is still debate about the functional forms that best describe these materials. One compilation\cite{zhou_low-frequency_2009} of data shows distinct behavior at low $f$, with $c(f)$ constant and $\alpha(f) \propto f^2$, and high $f$, with $c(f)$ increasing and $\alpha(f) \propto f^{1/2}$, with some crossover region characterized by a frequency $f = f^*$. This behavior is also typical of plane wave propagation in a viscous medium, and will be termed ``classical viscosity.'' These data can be fit to the Biot-Stoll model and its extensions~\cite{biot_theory_2005-1, biot_theory_2005, biot_generalized_2005,stoll_acoustic_1977, stoll_theoretical_1980, stoll_reflection_1981, stoll_marine_1985,chotiros2017Acoustics}, which treat the sediment as a porous medium in which the frame can support elasticity, and losses result from the relative motion between the fluid and the frame due to the viscosity of seawater. Its adaptation to sediments introduced losses within the poroelastic frame using a viscoelastic solid model\cite{stoll_marine_1985}. The majority of the Zhou, et al. \cite{zhou_low-frequency_2009}, data that show this dependence are based on indirect dispersion measurements, such as bottom reflection or waveguide propagation. It is also assumed that the sediment is a homogeneous half space. At those low frequencies, the acoustic waves interact with at least several meters of the sediment, which often contains gradients and layering structure. As recent work \cite{Godin_2021,godin_effects_2023} has shown, slight layer contrasts within a weakly shear supporting sediment may have significant impact on modal attenuation. Therefore, we focus on only direct measurements of sediment dispersion, either in situ, laboratory, or using an idealized sediment composed of glass beads.

\begin{figure}
    \centering
    \includegraphics[width=\columnwidth]{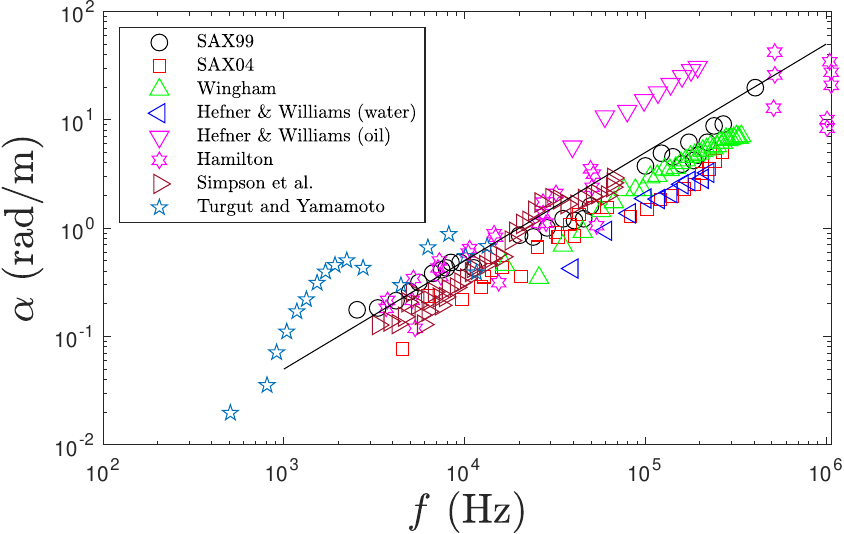}
    \caption{A compilation of direct measurements of the attenuation coefficient of saturated sand or glass beads. Field measurements from SAX99\cite{williams2002comparison}, SAX04\cite{Hefner2009}, Hamilton\cite{Hamilton1980}, Simpson et al.\cite{Simpson2003}, and Turgut and Yamamoto\cite{Turgut1990} are included here. The laboratory measurements are from  from Wingham\cite{Wingham1985} and Hefner and Williams\cite{hefner2006sound}. The solid black line shows $\alpha \propto f$. }
    \label{fig:lin-exp-data}
\end{figure}

Many of the direct measurements of dispersion in sandy sediments depart from classical viscosity, most notably showing an approximate $\alpha(f) \propto f$ relationship. The field studies cited here are from the measurements compiled by Hamilton \cite{Hamilton1980}, the large SAX99 \cite{williams2002comparison}, and SAX04 \cite{Hefner2009} experiments, as well as field measurements by Turgut and Yamamoto\cite{Turgut1990} and Simpson, et al.\cite{Simpson2003}. Laboratory data using sand are collected from Wingham \cite{Wingham1985}, and laboratory measurements using glass spheres were made by Hefner and Williams\cite{hefner2006sound} (using both water and oil as the saturating fluid). These measurements are collected in Fig.~\ref{fig:lin-exp-data}, and show a deviation from the Biot-Stoll model, and agreement with $\alpha \propto f$ for frequencies $1<f<400$~kHz.  We note that several experiments\cite{Lee2007,Sessarego2008} were excluded from our compilation of experimental data in Fig.~\ref{fig:lin-exp-data} due to the presence of negative velocity dispersion, an indication that the wavelength was close to the mean grain size of the medium (see Hare and Hay\cite{Hare2020} and references therein).

Trying to understand how such an attenuation form (specifically the $\alpha \propto f$ trend observed in SAX99\cite{Hamilton1980}) might arise was in part the motivation for Buckingham's grain-shearing (GS)~\cite{Buckingham1997,buckingham_wave_2000},  theory of sediment acoustics, and the later addition of viscosity~\cite{buckingham_pore-fluid_2007}. In his justification of this theory, Buckingham\cite{buckingham_wave_2000} noted that ``grains in an unconsolidated granular material do not form a macroscopic elastic skeletal frame,'' but instead that ``forces within the unconsolidated medium arise from grain-to-grain interactions'' which ``give rise to force chains.'' The term ``force chains'' refers to the fact that granular contact networks are spatially disordered, and the forces between grains are not homogeneous~\cite{majmudar2005contact}. 

Buckingham engaged with some of the pioneering work in the field of granular mechanics ~\cite{jaeger1996granular} to justify the need for a granular-focused approach. However, the key mechanisms in the GS theory were the frictional and time-dependent characteristics of the grain-grain contacts and not the spatial structure of the ``force chains'' explicitly. The time-dependent damping law used for grain-grain interactions was the key to producing $\alpha \propto f$. 

However, Buckingham's original works were published during the infancy of a revolution~\cite{forterre2008flows,van2009jamming,liu2010jamming,behringer2018physics} in our understanding of the mechanical properties of dense granular media, facilitated by improvements in experimental~\cite{amon2017preface} and simulation~\cite{radjai2017modeling,kamrin2024advances} methods and technologies. One key result is that disordered packings often have \textit{emergent} mechanical properties that are not \textit{a priori} obvious or consistent with a simple continuum picture. 

For example, disordered packings of elastic disks (2D) or spheres (3D) exhibit an excess of low-frequency vibrational modes~\cite{Silbert_PRL_2005,Wyart_PRE_2005}. In uniform elastic solids, low-frequency vibrations are assumed to be acoustic plane waves obeying Debye scaling, which implies a density of modes $\mathcal{D}(\omega) \propto \omega^2$ in three dimensions, where $\omega=2\pi f$ is the angular frequency. In contrast, the density of vibrational modes in disordered packings is empirically found to be very sensitive to the dimensionless pressure $\hat{P}$, defined as the ratio of system pressure to the bulk modulus of an individual grain. At large $\hat{P}$ (unrealistic for geological granular materials including marine sediments), the density of modes is consistent with the Debye picture, but these particles are ``soft'' and possess many more contacts than necessary. As $\hat{P}$ is decreased, the system develops an excess of low-frequency vibrational modes compared with Debye scaling~\cite{Silbert_PRL_2005}. 

How packing structure affects dispersion and attenuation in disordered packings is an open question, the answer to which would have obvious implications for marine sediment acoustics as well as other systems. We note a recent paper~\cite{saitoh2023sound}, which studied standing modes in disordered 2D packings. These authors demonstrated that, even at large $\hat{P}$, the disordered packing structure affects the dispersion relation and attenuation rate (in time) of attenuation of plane wave modes. However, many open questions remain, including the behavior at small $\hat{P}$, which is more appropriate to geological granular materials like marine sediments.

To this end, we here consider discrete-element method (DEM) simulations of granular packings and show that the granular packing structure plays a crucial role in determining the dispersion and attenuation curves. DEM simulations allow us to postulate grain-grain interaction rules and then solve the equations of motion for each grain individually. Thus, our approach explicitly considers the granular packing structure without assuming a continuum wave equation. This approach has limitations, which we discuss below in Sec.~\ref{sec:1D-theory}; however it is able to isolate the effect of the granular packing structure on sound speed and attenuation, which has not been explicitly considered previously.

We use linearized forces (linear springs and dashpots) between every grain, which is mathematically similar to viscous drag. The repulsive springs represent repulsive intergrain interactions, and the dashpots represent the combined effect of lossy grain-grain contacts and the effect of drag from the interstitial fluid (see Sec.~\ref{sec:1D-theory} for further discussion). Using these linearized forces we theoretically and computationally obtain the standard result of $\alpha\propto f^2$ at low $f$ and $\alpha \propto f^{1/2}$ at high $f$ in the continuum limit for a 1D periodic chain of spheres. However, in simulations of packings in two (2D) and three (3D) dimensions, our results show emergent scaling, where $\alpha \propto f^\beta$ at low $f$ with $\beta < 2$; the 3D simulations demonstrate that $\beta \approx 7/6$, which is very near 1. We find that 2D and 3D packings revert back to the continuum behavior at large $\hat{P}$, similar to the disappearance of excess low-frequency modes at large $\hat{P}$ discussed above.

Our results suggest that $\alpha \propto f^\beta$ with $\beta \approx 1$ scaling may be a direct result of the packing structure of granular materials, even with simple linear forces. The finding of $\alpha \propto f^1$ is consistent with many of the direct measurements of dispersion in sandy sediments\cite{williams2002comparison,hefner2006sound,Wingham1985,Hamilton1980}. Our results may be sensitive to other granular effects, including Hertzian grain interactions~\cite{daraio2005strongly,gomez2012shocks}, which are more realistic and involve a contact stiffness that increases with relative displacement. Other nonlinearities include contact breaking~\cite{owens_sound_2011,Schreck_PRL_2011}, frictional grain interactions~\cite{silbert2010jamming}. Non-spherical grain geometry~\cite{nguyen2014effect,nguyen2015effects} may also play a role. These effects may be considered in future studies. The problem of integrating the inertia and compressibility\cite{Williams2009b} of the interstitial fluid is key to explaining the behavior of in situ dispersion measurements and is another area for future work. However, the key result is that the granular packing structure plays a crucial role in determining the scaling laws for $c(f)$ and $\alpha(f)$ and therefore should be more explicitly included in theories of sediment acoustics.

The remainder of this paper is organized as follows. In Sec.~\ref{sec:1D-theory}, we consider the continuum limit of a 1D granular model. In Sec.~\ref{sec:sim-methods}, we describe the numerical simulations we use to compare to this continuum theory. Results of the numerical simulations are shown in Sec.~\ref{sec:results}, where we demonstrate how anomalous scaling laws emerge in 2D and 3D due to the granular packing structure. In Sec.~\ref{sec:discussion} we provide discussion and concluding remarks.

\section{1D linear continuum theory}
\label{sec:1D-theory}
\begin{figure}[h]
\begin{center}
\includegraphics[width=\columnwidth]{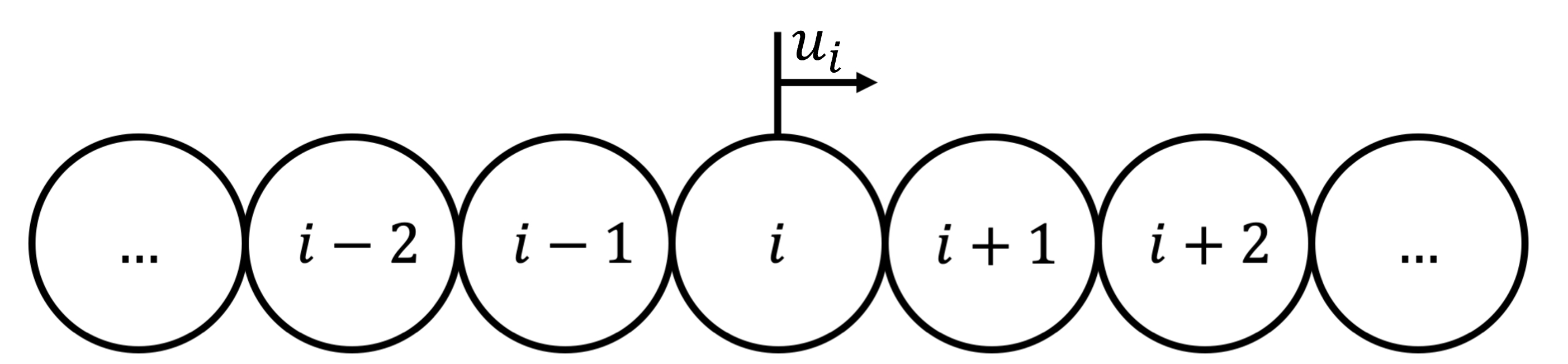}
\end{center}
\caption{A one dimensional chain of identical spherical grains.}
\label{fig:GrainChain} 
\end{figure}

We first consider the continuum limit of a 1D granular model, based on the dynamics of a single ``force chain'' of grains, as shown in Fig.~\ref{fig:GrainChain}. We use this theory as a baseline to compare numerical simulations in 1D, 2D, and 3D. Each grain has identical diameter $d$ and mass $m$. We assume grains $i$ and $i+1$ interact via dissipative, repulsive contact forces. These forces are linearized by assuming that the repulsive forces use linear springs and the dissipative forces use simple dashpots. Physical grains have nonlinear repulsive forces between them (e.g. the Hertz contact law~\cite{coste1999validity}), but a linear approximation is reasonable for small displacements such as acoustic strains. 

One limitation of this approach is that the grains must be considered as perfectly rigid, and compression of grains is modeled by the finite overlap between them. This approximation is only valid in the limit where time scales associated with dynamics inside the grains (in this case, internal elastic waves) are very small compared with other dynamical time scales in the system (e.g., shear rates or, in this case, propagation of waves in the bulk material). Marine sediments are on the edge of validity of such approximations, since the sound speed in, e.g., silica grains is only about four times that of water. Future work should more closely examine if and how treating the grains as objects with internal elasticity (e.g., an FEM approach) affects the results.

The linear dashpot term assumes that the dissipation is directly proportional to the relative velocity between the grains at any given time. Such a functional form is meant to represent some combination of effects from the interstitial fluid as well as inelastic effects that are inherent to grain-grain contacts.

With these assumptions, the displacement $u_i$ of grain $i$ from its unperturbed position $x$ on the chain is then governed by the following differential equation, which includes a repulsive spring force and a dashpot for each neighbor $i-1$ at position $x-d$ and $i+1$ at position $x+d$:
\begin{align}
    m \ddot{u}_i = \kappa& (u_{i-1}-u_i) - \kappa (u_i -u_{i+1}) \nonumber \\
    & + \gamma (\dot{u}_{i-1} - \dot{u}_i) - \gamma (\dot{u}_i - \dot{u}_{i+1})\, ,
    \label{eqn:1D-force-law}
\end{align}
Here, $\kappa$ is a spring constant, $\gamma$ is the dashpot coefficient, and dots denote time derivatives. Taking the long-wavelength limit (i.e., assuming $d$ is small enough to transform the finite differences into derivatives) yields
\begin{equation}
    mu_{tt} = \kappa d^2u_{xx} + \gamma d^2u_{txx},
    \label{eqn:1D-wave-eq}
\end{equation}
where subscripts $t$ and $x$ now denote derivatives with respect to time and space, respectively. A decaying plane wave is guessed, given by
\begin{equation}
    u(x, t) = Ae^{-\alpha x}e^{i(kx - \omega t)},
    \label{eqn:deacying_plane_wave}
\end{equation}
where $\omega$ is the angular frequency, $k$ is the acoustic wavenumber, and $\alpha$ is the attenuation parameter with units of radians per meter.
We can then solve the system to arrive at the two following equations for non-dimensional wavenumber and attenuation:
\begin{eqnarray}
     \hat{\alpha}&=& \hat{\omega}^2 \frac{\hat{\gamma}}{\sqrt{2}}\left\{ \left[1+(\hat{\omega} \hat{\gamma} )^2 \right] \left[1+\sqrt{1+(\hat{\omega} \hat{\gamma} )^2}\right] \right\}^{-1/2}      \label{eqn:nondim-absorption} \\
     \hat{k}&=& \hat{\omega} \frac{1}{\sqrt{2}}\frac{\left\{ \left[1+(\hat{\omega} \hat{\gamma} )^2 \right] \left[1+\sqrt{1+(\hat{\omega} \hat{\gamma} )^2}\right] \right\}^{1/2}}{1+(\hat{\omega} \hat{\gamma} )^2 }\, .
     \label{eqn:nondim-dispersion}
\end{eqnarray}
The dimensionless quantities in Eqs.~\eqref{eqn:nondim-absorption} and \eqref{eqn:nondim-dispersion} are defined by the following equations:
\begin{align}
     \hat{k} &= kd \\
     \hat{\alpha} &= \alpha d \label{eqn:alphahat}\\
     \hat{\omega} &= \omega \sqrt{m/\kappa} \label{eqn:omegahat}\\
     \hat{\gamma} &= \gamma/ \sqrt{\kappa m} \label{eqn:gammahat}\, .
\end{align}
The wavenumber and attenuation coefficient each have dimensions of inverse meters, and are easily non-dimensionalized by the grain diameter. Angular frequency is non-dimensionalized by the resonance angular frequency for an individual grain contact, and the dashpot term is non-dimensionalized using the ratio of the energy lost and energy stored per cycle, which is related to the Q factor of the contact \cite{Garrett2020a}.

We note that Eq.~\eqref{eqn:nondim-absorption} gives $\alpha\propto \omega^2$ for low frequencies and $\alpha \propto \omega^{1/2}$ at high frequencies. For wave speed, the result is constant at low frequencies and then increases according to $c\propto \omega^{1/2}$ at high frequencies. Equation~\eqref{eqn:1D-wave-eq} has a similar mathematical form to classic viscosity \cite{shitikova2022models}, so these results are expected.

\section{Methods}
\label{sec:sim-methods}
We now discuss the methods used to perform DEM numerical simulations in 1D, 2D, and 3D of granular packings subjected to harmonic driving at a boundary.

\subsection{Numerical Integration Scheme}
\label{sec:methods-num-int}
For each grain $i$, we use a modified velocity-Verlet~\cite{binder2004molecular} scheme,
\begin{align}
\mathbf{r}(t+dt) &= \mathbf{r}(t)+\mathbf{v}(t)dt+\frac{1}{2}\mathbf{a}(t)dt^2 \\
\mathbf{v}(t+dt) &= \mathbf{v}(t)+\frac{1}{2}[\mathbf{a}(t)+\mathbf{a}(t-dt)]dt
\end{align}
to integrate Newton's equations of motion. At each time step, the acceleration for each grain $i$ is calculated using
\begin{equation}
    m_i \mathbf{a}_i = \sum_j \mathbf{F}_c^{ij} + \sum_j \mathbf{F}_d^{ij}
    \label{eqn:ma=Fc+Fd}
\end{equation}
Here, $m_i$ is the mass of grain $i$; $\mathbf{r}_i$, $\mathbf{v}_i$, and $\mathbf{a}_i$ are the vector position, velocity, and acceleration of grain $i$; $\mathbf{F}_c^{ij}$ are repulsive contact forces between grains $i$ and $j$ that depend on their positions; $\mathbf{F}_d^{ij}$ are dissipative forces that depend on the velocities of grains $i$ and $j$; and the sum over $j$ denotes a sum over all grains $j$ that are in contact with grain $i$.

\subsection{Intergrain forces}
\label{sec:methods-forces}
We use linearized repulsive forces that involve overlaps $\xi_{ij} = ( R_i + R_j - |\mathbf{r}_i - \mathbf{r}_j| )\Theta(R_i + R_j - |\mathbf{r}_i - \mathbf{r}_j|)$. Here, $R_i$ and $R_j$ are the radii of particles $i$ and $j$ and $\Theta$ is the Heaviside function, which sets $\xi$ to zero when particles are not in contact. The repulsive force is modeled in terms of this overlap,
\begin{equation}
    F_c^{ij} = -\kappa \xi_{ij} \hat{\mathbf{n}}_{ij}\, ,
    \label{eqn:contact-force}
\end{equation}
where the normal vector $\hat{\mathbf{n}}_{ij}$ has magnitude of 1 and points along the line from the center of particle $i$ to the center of particle $j$. We note that the amplitudes involved in acoustic waves are sufficiently small that individual grain-grain contacts are well approximated by linear repulsive forces.

We also use linearized dissipative forces using a damping coefficient $\gamma$. For the bulk of the results we show, the dissipative forces can be simply written as
\begin{equation}
    F_d^{ij} = -\gamma (\mathbf{v}_i - \mathbf{v}_j).
    \label{eqn:damping-force}
\end{equation}
We note that this form damps all relative motion between grains, both normal and tangential. Such a form is intended to be the simplest form that mimics a combination of lossy grain-grain contacts combined with fluid-mediated damping due to relative motion in saturated granular packings.

The functional form in Eq.~\eqref{eqn:damping-force} assumes that normal and tangential motion are damped with the same magnitude (note that this only applies in 2D and 3D, since there is no tangential motion in 1D). However, even if a linear approximation is reasonable, there are likely differences in the magnitude of damping for normal and tangential relative motion between grains. In this case, we can write in 2D, using $\delta \mathbf{v}_{ij} \equiv (\mathbf{v}_i - \mathbf{v}_j)$,
\begin{eqnarray}
    F_d^{ij} =& -\gamma_n \hat{\mathbf{n}}_{ij}( \delta \mathbf{v}_{ij} \cdot \hat{\mathbf{n}}_{ij})
    -\gamma_t \hat{\mathbf{s}}_{ij} (\delta \mathbf{v}_{ij}\cdot \hat{\mathbf{s}}_{ij}).
    \label{eqn:damping-force-decomp}
\end{eqnarray}
Here, $\mathbf{s}_{ij}$ is a tangential unit vector such that $\mathbf{s}_{ij}\cdot \mathbf{n}_{ij}=0$. If $\gamma_n = \gamma_t$, this form reduces to the form given in Eq.~\eqref{eqn:damping-force}. While we primarily use Eq.~\eqref{eqn:damping-force}, we also show simulations in 2D using Eq.~\eqref{eqn:damping-force-decomp} with $\gamma_t = 0$ to support our claims in Sec.~\ref{sec:2D-results}. In 3D, two tangential unit vectors would be required.

We also note that use of $\gamma_n$ and $\gamma_t$ is common in DEM simulations of dry granular media (i.e., with no interstitial fluid). This is because grain-grain collisions involve restitution losses, i.e., the conversion of some of kinetic energy of the grains into heat. The magnitude of these losses is often parameterized by a restitution coefficient $e_n$, which can be connected to the magnitude of $\gamma_n$ for perfectly normal collisions. In particular, for linear interactions~\cite{force_schemes}, 
\begin{equation}
    e_n = \exp\left(-\frac{\gamma_n}{2m_{\rm eff}} t_n \right),
\end{equation}
with
\begin{equation}
    t_n = \pi \left[ \frac{\kappa}{m_{\rm eff}} - \left(\frac{\gamma_n}{2m_{\rm eff}} \right)^2 \right] ^{-1/2}
\end{equation}
where $m_{\rm eff}$ is the reduced mass.
These normal losses have not been considered by the acoustics community to date, although the work of Buckingham\cite{Buckingham1997} focused on modeling of a time-dependent $\gamma_t$-like term. Some evidence for grain-based losses is also provided by Hefner and Williams\cite{hefner2006sound} who found that increasing the kinematic viscosity of the pore fluid by a factor of 100 only increased the attenuation coefficient at the same frequency by a factor of ten. This situation is possible if the effective $\gamma$ from grain contacts is ten times that of viscous drag from water, and one tenth that of viscous drag from silicone oil. Given estimates of $e_n$ from glass beads of approximately 0.98 for small velocities\cite{Montaine2011} the resulting $\gamma_n$ results in the right order of magnitude to explain the results of Hefner and Wiliams\cite{hefner2006sound}.

Thus, a linearized dashpot, which applies a force proportional to the magnitude of relative velocity between particles, is a reasonable model to use both for viscous-like forces as well as inherently lossy grain-grain contacts. Our $\gamma$ parameters are meant to capture some combination of these effects. Using $\gamma_n$ to control $e_n$ models losses during collisions, not necessarily due to relative motion during harmonic oscillation of a packing, as we do here. In the latter case, the contacts are enduring. Future work is needed to better understand the relative contributions of losses due to viscous effects and lossy contacts.

\subsection{Generating Packings at Different Pressures}
\label{sec:methods-packings}
In 1D, initial configurations are trivial to generate, as they constitute a periodic chain. However, in 2D and 3D, grains must be prepared in a disordered packing with all grains in force balance. Here we describe how disordered packings are generated using DEM simulations, and we discuss both 2D and 3D situations at the same time, calling out differences as appropriate. Unless otherwise stated, the discussion below applies to both 2D and 3D packing generation protocols.

Disorder is enforced using two diameters of grains with ratio 1.4, with equal number of particles of each size~\cite{ohern2003jamming}. Using two grain sizes ensures that 2D packings do not end up with a periodic, crystaline structure. We consider $N$ grains, where $N$ varies between 5,000 and 20,000, half with diameter $d$ and half with a larger diameter $1.4d$. Despite having different diameters, all grain masses are set to the same value $m$. Grains are initially placed randomly throughout a large domain with fixed walls in the $x$-direction at $x=0$ and $x=L_x$ and periodic boundaries in the $y$ and $z$ directions with length $L_y$ and $L_z$ (in 3D only). The periodic boundaries simulate an infinite medium in the transverse directions, while a nonperiodic (wall-based) boundary is required to observe the propagating and decaying wave as it moves down the channel.

During the computationally intensive initial compression, we also include a dissipative force that is linearly dependent on the absolute velocity of the grain, as if the grains were in a highly viscous background fluid. Additionally, we set the velocity and acceleration to zero for any particle with no contacts, which significantly speeds up the compression phase. These additional damping mechanisms are turned off after the initial packings are generated.

Compression of the system is done in the following way. Grains are initially placed randomly, without touching, with very large $L_y \approx L_x$. The value of $L_x$ is held fixed throughout compression, typically between $200 < L_x/d < 1000$.  Additionally, in 3D, $L_z$ is held fixed throughout the simulation, typically between $5 \leq L_z/d \leq 10$. At each compression step, we compress the $y$-direction using the substitution $L_y\rightarrow L_y (1-\Delta)$ as well as substituting the $y$-position of each particle using $y_i \rightarrow y_i (1-\Delta)$, where $\Delta$ is chosen to achieve the desired $\hat{P}$ (typically, $\Delta \approx \hat{P}$. After such a compression step, we allow the particles to rearrange and the energy to dissipate by evolving the simulation in time. When the total kinetic energy becomes sufficiently small, we compress again and again allow the system to relax. During compression, we track the mean kinetic energy per particle $\bar{K} = \frac{1}{N}\sum_i \frac{1}{2}m v_i^2$ and potential energy per particle $\bar{U} = \frac{1}{2N} \sum_{i} \sum_{j} \frac{1}{2}\kappa \xi_{ij}^2$, where the extra factor of 2 in the denominator of $\bar{U}$ accounts for double counting of contacts.

The system pressure is defined $P = \frac{1}{d^{D-1}} \sqrt{2\bar{U}\kappa}$, where $D$ is the number of spatial dimensions. We use the dimensionless pressure $\hat{P}$ to characterize the packings, with $\hat{P} = P/\kappa$ in 2D and $\hat{P} = Pd/\kappa$ in 3D, i.e.,
\begin{equation}
    \hat{P} = \sqrt{\frac{2\bar{U}}{\kappa d^2}}.
    \label{eqn:P-hat}
\end{equation}
Physically, $\hat{P}$ can also be thought of as an approximation of the characteristic size of the overlap between two particles compared to a particle size, $\bar{\xi}/d$. Realistic values of $\hat{P}$ for marine sediments may be extremely small, since confining pressures are generated by the weight of grains above them, e.g., $\rho g h \approx 25$~kPa for $\rho = 2500$~kg/m$^3$, $g = 9.81$ m/s$^2$, and $h = 1$~m. The stiffness of the grains is tens of GPa, so $\hat{P} \approx 10^{-6}$ may be a physically reasonable value. Although packings at such low $\hat{P}$ are difficult to generate computationally, it is often true that there is a low-pressure limit at a value of $\hat{P}$ that is much higher and therefore much easier to generate. In this low-pressure limit, the dynamics are insensitive to the value of $\hat{P}$ so long as it is sufficiently low.

\begin{figure}
    \raggedright
    (a) \\
    \includegraphics[width=\columnwidth]{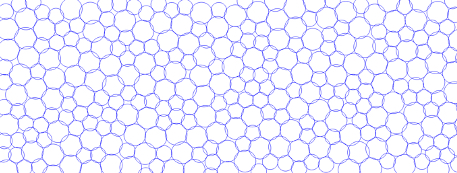}\\
    (b) \\
    \includegraphics[width=\columnwidth]{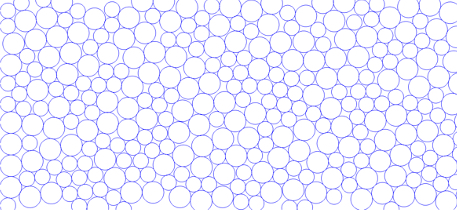}
    \caption{Two packings, one at $\hat{P} \approx 0.1$ (a) and another at $\hat{P} \approx 0.001$ (b).}
    \label{fig:packing-pics-2D}
\end{figure}

Thus, our approach here is to prepare packings at a range of values of $\hat{P}$ spanning $10^{-1}$ down to $10^{-4}$ until a low-$\hat{P}$ limit is reached. Despite the fact that the large-$\hat{P}$ packings are unrealistic, our approach has two major benefits. First, it allows us to show that the low-pressure limit occurs for our simulations at $\hat{P} \approx 0.005$, and further decreasing $\hat{P}$ has no effect on the results. Second, our approach allows us to show that the acoustic properties of granular packings agree with the 1D continuum theory at large $\hat{P}$, whereas anomalous scaling appears as the low-pressure limit is approached. Packings in 2D at high and low $\hat{P}$ are shown in Fig.~\ref{fig:packing-pics-2D}. The high $\hat{P}$ packing is far from the unjamming transition~\cite{Silbert_PRL_2005,van2009jamming}, where the packing would break apart and have no intergrain forces, which can be seen from its large grain-grain overlaps and relatively large number of contacts per grain. It is thus expected to behave more like a continuum elastic solid than the low $\hat{P}$ packing.

\section{Results}
\label{sec:results}
\subsection{1D Simulations \label{sec:1D-results}}
We first perform simulations in 1D. These simulations both confirm our theoretical analysis above as well as demonstrate that the effects observed in higher dimensions are not an artifact of DEM simulations but a result of disordered granular mechanics in higher spatial dimension.

The initial positions of grains are prepared, in 1D, by assigning the grains' positions at equally spaced positions similar to Fig.~\ref{fig:GrainChain}. In 1D, $\mathbf{r}_i = x_i \hat{\mathbf{x}}$, where $\hat{\mathbf{x}}$ is a unit vector along the chain. We study 10,000 equal sized grains placed in equal spacing $d = 1$ along a line. Each grain is subject to forces from two neighbors, according to Eq.~\eqref{eqn:1D-force-law}. The first grain in the chain is driven in one of two ways. The first method, constant driving, begins driving at $t=0$ according to $x_1(t) = x_1(0)+A \sin(\omega t)$. We choose $A/d = 10^{-4}$ and have verified that our results are not sensitive to this choice. Second, we drive the grain according to a Gaussian-enveloped pulse with a primary angular frequency of $\omega$ and a bandwidth of $\omega/5$. These two types of driving yield similar results.

We measure $\alpha$ for a given $\omega$ by measuring the peak height of oscillation of each grain as a function of its distance down the chain and fitting to a decaying exponential. For the constant-driving case, we measure the wavenumber $k$ by observing the phase $\phi$ of each grain and fitting to $\phi = kx$. This allows us to infer $c = \omega / k$. For the Gaussian-enveloped pulse, we measure $c$ using the time corresponding to the largest displacement of each grain. The results from these two methods are indistinguishable.

\begin{figure}
\includegraphics[width=\columnwidth]{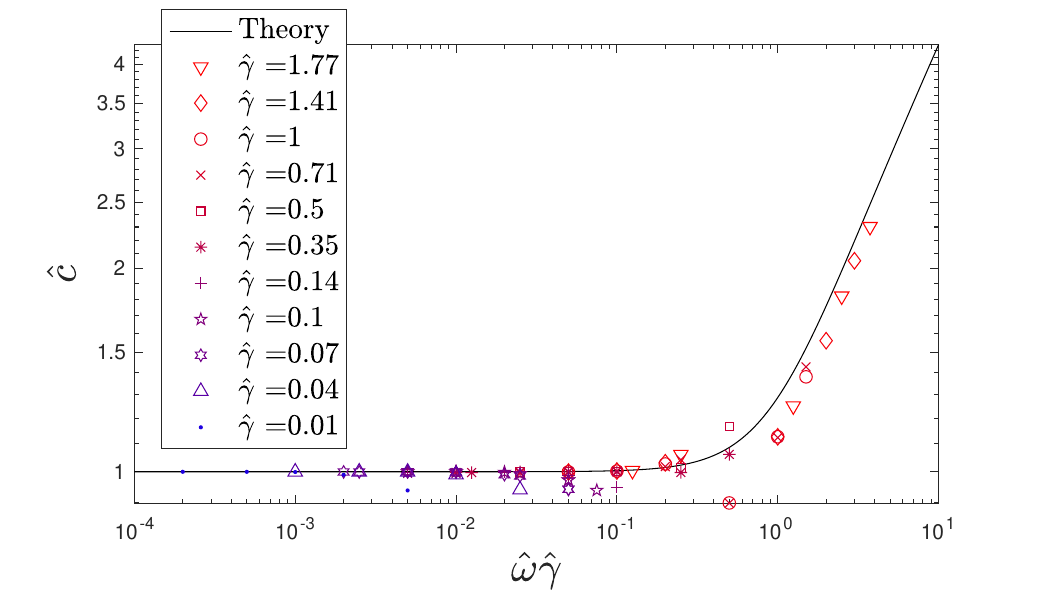}
\includegraphics[width=\columnwidth]{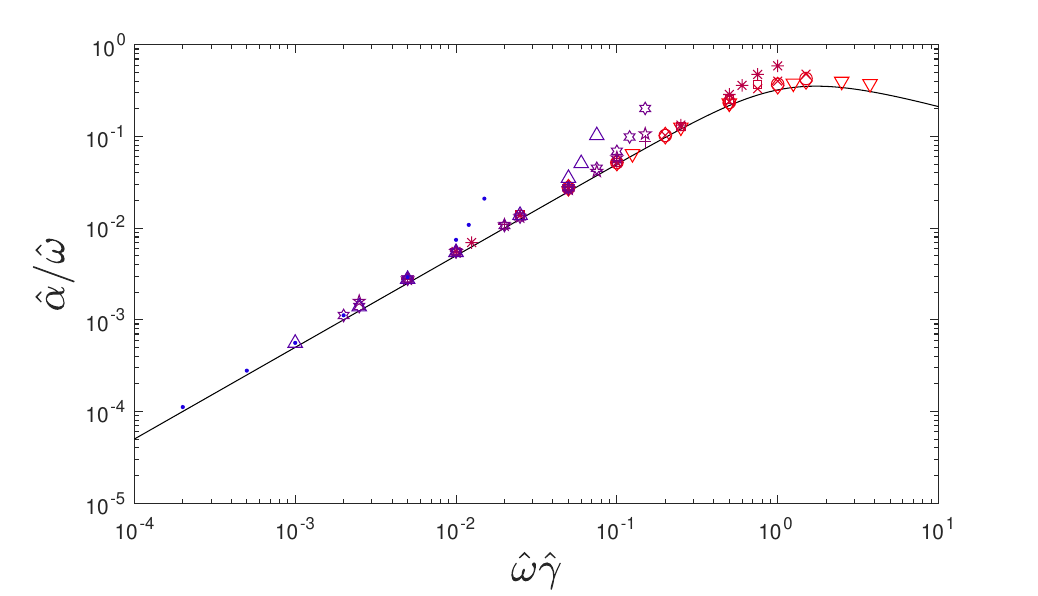}
\caption{Results for 1D simulations. Dimensionless wave speed, $\hat{c}$, and ratio of attenuation to frequency, $\hat{\alpha}/\hat{\omega}$, both plotted as a function of dimensionless frequency times dimensionless damping, $\hat{\omega}\hat{\gamma}$. Solid black lines show the theoretical result from Eqs.~\eqref{eqn:nondim-absorption} and \eqref{eqn:nondim-dispersion}.}
\label{fig:1D-results}
\end{figure}

Results of these measurements are shown in Fig.~\ref{fig:1D-results}. We systematically vary $\kappa$, $\omega$, and $\gamma$ in order to access a wide range of values of $\hat{\omega} \hat{\gamma}$. The measured data from simulations agree well with the continuum theory from Sec.~\ref{sec:1D-theory} over a large range of $\hat{\gamma}$.

The fact that the 1D simulations do not significantly deviate from the 1D theory is not surprising, since the theory was developed to model this exact situation. These results are presented as an example to provide confidence that the observed emergent scaling behavior in 2D and 3D is not an artifact of the simulation methods but instead comes from the granular packing structure.

\subsection{2D Simulations}
\label{sec:2D-results}

In 2D and 3D, we perform similar simulations. However, the initial configurations of grains are not along a line but are generated by the packing procedure described above in Sec.~\ref{sec:methods-packings}. Before describing the results of the phase speed and attenuation measurements, we first discuss some features of the packings themselves.

\begin{figure}
    \raggedright
    (a) \\ \centering
    \includegraphics[width=\columnwidth]{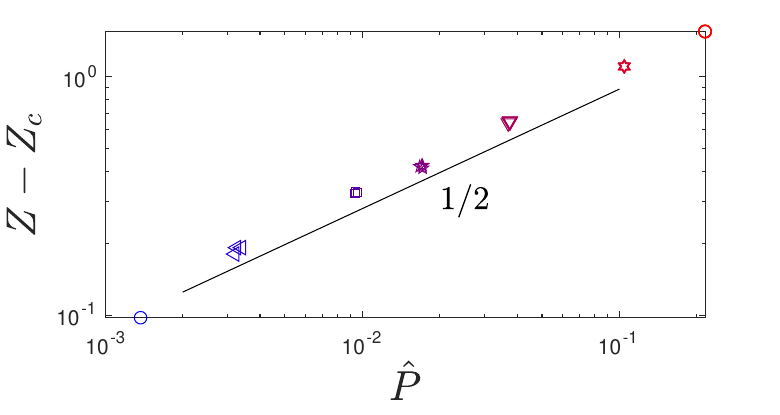} \\
    \raggedright
    (b) \\ \centering
    \includegraphics[width=\columnwidth]{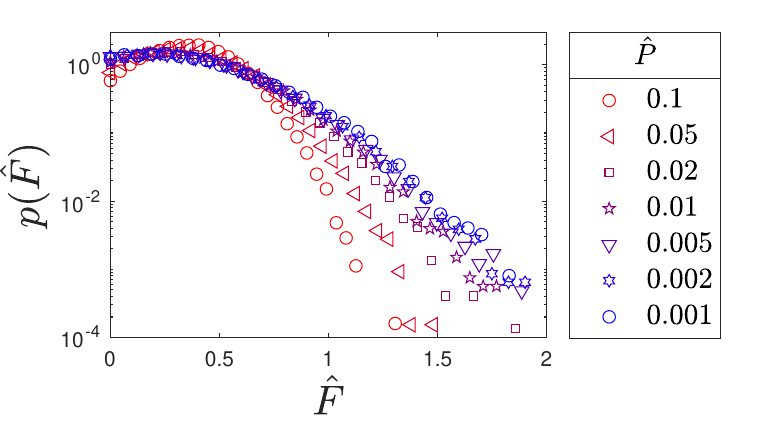}
    \caption{(a) Excess contacts $Z-Z_c$ plotted as a function of dimensionless pressure $\hat{P}$, showing $Z-Z_c \propto \hat{P}^{1/2}$ (solid line). (b) The probability distribution $p(\hat{F})$ of dimensionless contact forces $\hat{F}$. Changing colors (blue to red) represent increasing $\hat{P}$, with the same color and symbol convension between panels (a) and (b).} 
    \label{fig:packing-stats-2D}
\end{figure}

A common way to characterize granular packings involves characterizing the number of contacts per particle $Z$ using the concept of isostaticity~\cite{van2009jamming}. When a system is isostatic, the number of degrees of freedom is equal to the number of constraints. For $N$ frictionless particles in 2D, there are $2N$ degrees of freedom. Thus, isostaticity for frictionless disks in 2D requires $2N$ total grain-grain contacts, or $Z = 4$ contacts per particle, since each contact is shared between two grains. This relation is valid for large systems where constraints related to the boundaries are not important. The isostatic limit is denoted $Z_c$. Figure~\ref{fig:packing-stats-2D}(a) shows that $\Delta Z \equiv Z-Z_c \propto \hat P^{1/2}$, meaning that $Z \rightarrow Z_c$, with $Z_c = 4$ in this case, as $\hat{P}\rightarrow 0$; this agrees with many previous studies~\cite{ohern2003jamming,van2009jamming,thompson2019critical}. We find similar results for 3D packings, but with $Z_c = 6$ due to the extra degree of freedom for each particle.

For sufficiently low $\hat{P}$ and $\Delta Z$, the statistical properties of the granular force network become nearly independent of pressure. For example, Fig.~\ref{fig:packing-stats-2D}(b) shows a probability distribution $p(\hat{F})$ of dimensionless contact force $\hat{F} \equiv F_c / P d$ (note that in 2D pressure has units of force per length). These force distributions show exponential tails with some dependence on pressure. However, the distributions become independent of $\hat{P}$ for $\hat{P}\leq 0.005$. For the high-pressure packings, which have large overlaps between particles, there are extra contacts in the force network leading to a narrower distribution of $\hat{F}$. These packings with large $\hat{P}$ are certainly not physically similar to marine sediments, due to large particle deformations, and are also unphysical in the sense that real grains would deform and fracture at such high strains.

\begin{figure}
    \raggedright (a) \\
    \centering
    \includegraphics[width = \columnwidth]{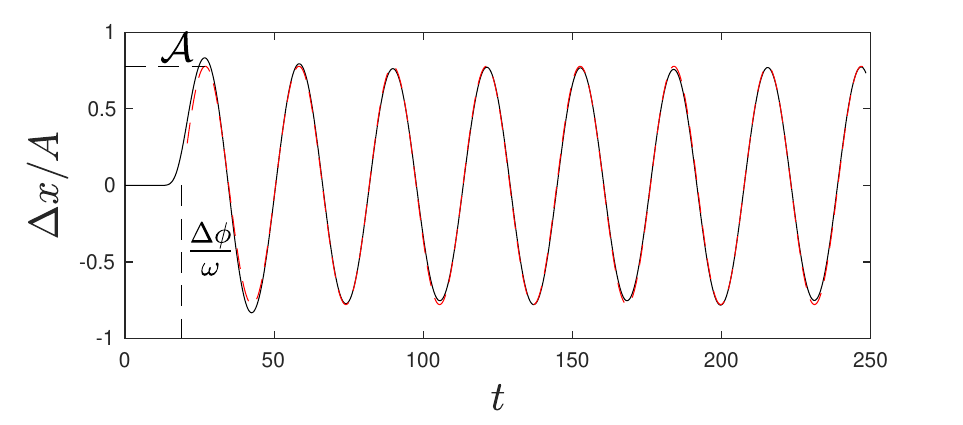}\\
    \raggedright (b) \hspace{70mm} (c) \\
    \centering
    \includegraphics[width = 0.49\columnwidth]{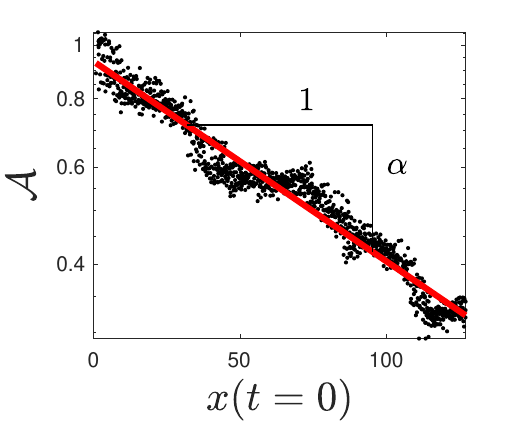}
    \includegraphics[width = 0.49\columnwidth]{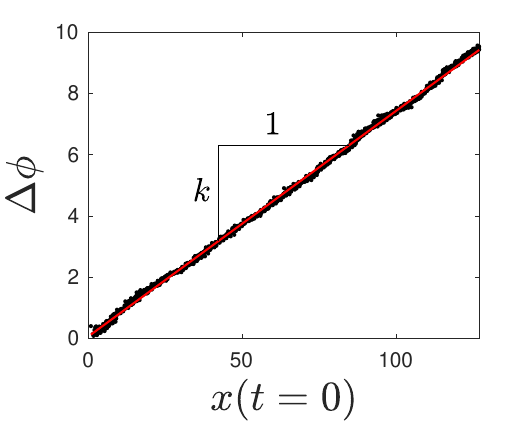} \\
    \caption{(a) Normalized displacement of an individual grain, $\Delta x/A$ (solid black curve) is plotted as a function of time, which can be fit to an sinusoid of the form $\mathcal{A} \sin (\omega t - \Delta \phi)$ (dashed red curve). (b,c) The constants $\mathcal{A}$ (b) and $\Delta \phi$ (c) can then be plotted as a function of the initial $x$ position of each grain and then used to extract the wavenumber $k$ and the attenuation coefficient $\alpha$.}
    \label{fig:2D-single-grain-dyn}
\end{figure}

Using these packings, we vibrate the wall at $x=0$ and measure the resulting displacements of all grains as a function of time. A plot of the $x$ displacement of a single grain is shown in Fig.~\ref{fig:2D-single-grain-dyn}(a). For every grain, we fit to a form $\Delta x/A = \mathcal{A} \sin (\omega t - \Delta \phi)$ and extract the normalized oscillation amplitude $\mathcal{A}$ and the phase offset $\Delta \phi$, as shown in the figure. These are plotted as a function of the $x$ position along the channel in panels (b) and (c) of Fig.~\ref{fig:2D-single-grain-dyn}. The slope of $\Delta \phi$ versus $x(t=0)$ gives us the wavenumber $k$, and the magnitude of the slope of $\log \mathcal{A}$ versus $x(t=0)$ gives us $\alpha$. We perform such measurements for multiple packings per pressure, widely varying $\omega$, $\gamma$, $\kappa$, and $m$. We then can obtain the dimensionless quantities $\hat{\alpha}$ and $\hat{k}$ as a function of the input parameters $\hat{\omega}$ and $\hat{\gamma}$ to test the agreement with the continuum theory.

\begin{figure}
    \raggedright
    (a) \\
    \includegraphics[width=\columnwidth]{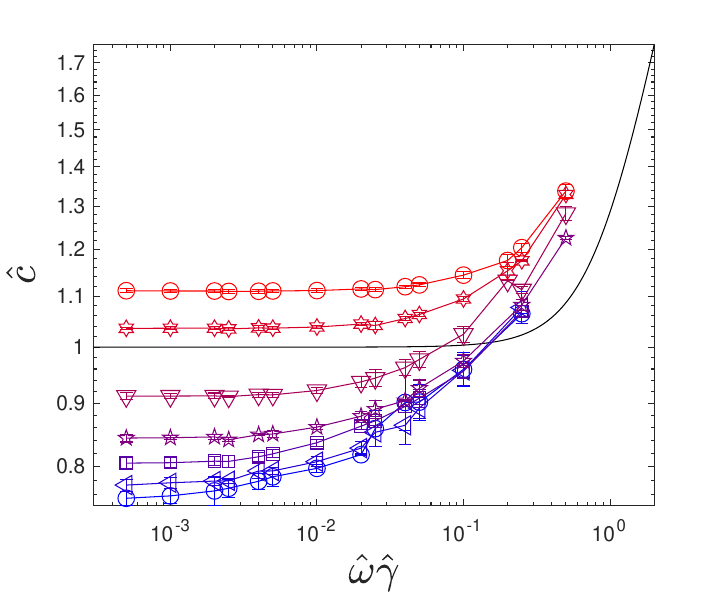} \\
    (b) \\
    \includegraphics[width=\columnwidth]{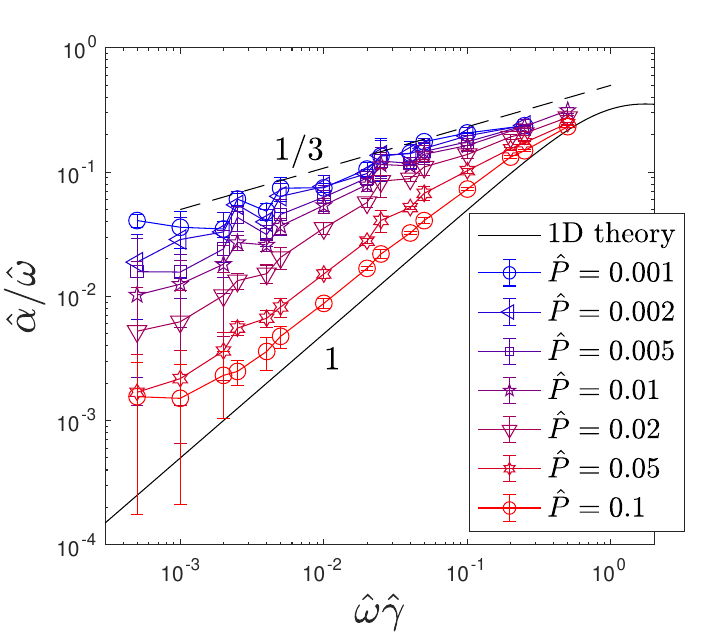}
    \caption{Dispersion and attenuation in 2D with varied pressure for compression waves. Plots show $\hat{c}$ and $\hat{\alpha}/\hat{\omega}$ versus $\hat{\omega}\hat{\gamma}$. Solid black lines show the theoretical result from Eqs.~\eqref{eqn:nondim-absorption} and \eqref{eqn:nondim-dispersion}.}
    \label{fig:2D-results}
\end{figure}

Figure~\ref{fig:2D-results} shows the key result of this paper, which are measurements of sound speed and attenuation for the packings characterized in Fig.~\ref{fig:packing-stats-2D}, using only linear springs and dashpots at each contact. In accordance with the 1D continuum theory in Sec.~\ref{sec:1D-theory}, we plot $\hat{c}$ and $\hat{\alpha}/\hat{\omega}$ as a function of dimensionless frequency $\hat{\omega} \hat{\gamma}$ for packings with pressures ranging from $\hat{P}=0.001$ to 0.1. Each data point represents an average measurement over multiple realizations of packings at the same $\hat{P}$ as well as multiple combinations of parameter values to form the dimensionless quantity $\hat{\omega}\hat{\gamma}$. We do this to check that the particular combinations of dimensionless quantities being plotted are still valid (despite the disagreement with the continuum theory from which they follow). For this reason, we show error bars using the standard deviation of the measurements being averaged (not the standard error, which would be the same quantity divided by the square root of the number of measurements). Most of the error bars are smaller than the symbols themselves, except for the measurements of very small $\alpha$ (since very little attenuation occurs over the entire channel) and large $c$ (since these are associated with very high driving frequencies and large attenuation). This shows that for multiple parameter combinations, $\hat{c}$ and $\hat{\alpha}/\hat{\omega}$ versus $\hat{\omega}\hat{\gamma}$ is still able to collapse the data, albeit with different scaling forms for different values of $\hat{P}$. 

The key message from these plots is that the functional forms of $c(f)$ and $\alpha(f)$ for the low-pressure (i.e., more realistic) packings do not agree with the continuum picture but instead follow a new, emergent functional form. For large $\hat{P}$, we recover scaling functions for $\hat{c}$ and $\hat{\alpha}/\hat{\omega}$ as a function of $\hat{\omega}\hat{\gamma}$ that agree with the continuum picture (solid black lines), e.g., $\hat{\alpha}/\hat{\omega} \propto \hat{\omega}\hat{\gamma}$ at low frequencies. However, as $\hat{P}$ is decreased, we observe that $\hat{c}$ and $\hat{\alpha}/\hat{\omega}$ asymptotically approach new scaling forms. In particular, $\hat{\alpha}/\hat{\omega} \propto (\hat{\omega}\hat{\gamma})^{1/3}$, which implies $\alpha \propto f^{4/3}$ for small $\hat{\omega}\hat{\gamma}$ in 2D. Additionally, rather than $\hat{c}$ constant for $\hat{\omega}\hat{\gamma} < 0.1$, $\hat{c}$ weakly increases over $10^{-3} < \hat{\omega}\hat{\gamma} < 10^{-1}$ before joining the high-pressure curves at $\hat{\omega}\hat{\gamma} \approx 0.5$ and above. In measurements of sediment dispersion \cite{williams2002comparison}, there typically exists a finite low-frequency asymptote, which is related to the inertial effects of the saturating fluid, as discussed by Williams\cite{Williams2009b}. Further model development using a saturating fluid is key for explaining the low frequency sound speed behavior.

\begin{figure}
    \raggedright
    (a) \\
    \includegraphics[width=\columnwidth]{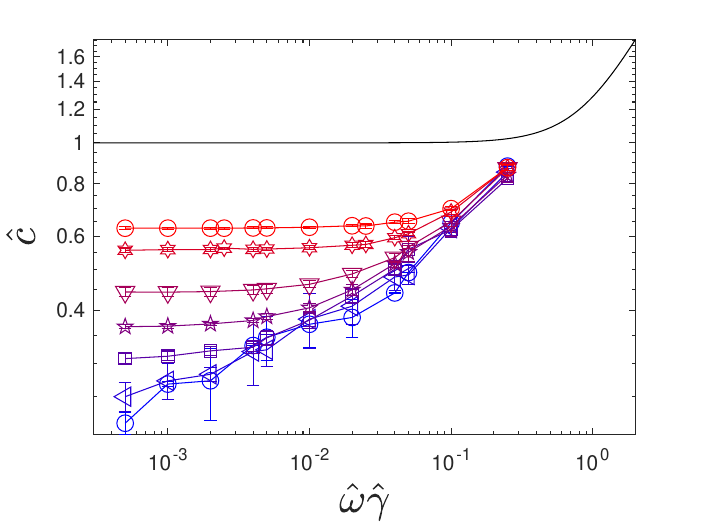} \\
    (b) \\
    \includegraphics[width=\columnwidth]{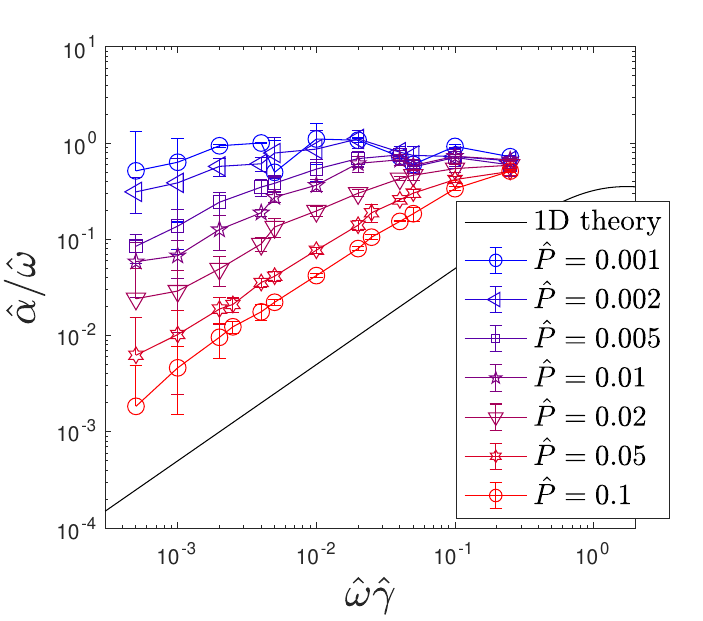}
    \caption{Dispersion and attenuation in 2D with varied pressure for shear waves. Plots show $\hat{c}$ and $\hat{\alpha}/\hat{\omega}$ versus $\hat{\omega}\hat{\gamma}$.  Solid black lines show the theoretical result from Eqs.~\eqref{eqn:nondim-absorption} and \eqref{eqn:nondim-dispersion}.}
    \label{fig:2D-results-shear}
\end{figure}

Figure~\ref{fig:2D-results-shear} shows similar measurements but with shear waves instead of compressional waves. For these measurements, we vibrate the wall as before but in the orthogonal lateral direction, tracking the lateral motion of particles down the channel to obtain $\alpha$ and $k$. We then plot the same dimensionless combinations as before. At high $\hat{P}$, the curves for speed and attenuation mimic the continuum theory, albeit shifted in both magnitude as well as the crossover value of $\hat{\omega}\hat{\gamma}$. As $\hat{P}$ is decreased, the curves shift to lower $\hat{c}$ with less of a plateau at low $\hat{\omega}\hat{\gamma}$ as well as an emergent scaling of $\hat{\alpha}/\hat{\omega}$ roughly constant for small $\hat{\omega}\hat{\gamma}$. This implies that $\alpha \propto f$ for shear waves in this system.

\begin{figure}
    \raggedright
    (a) \\
    \includegraphics[width=\columnwidth]{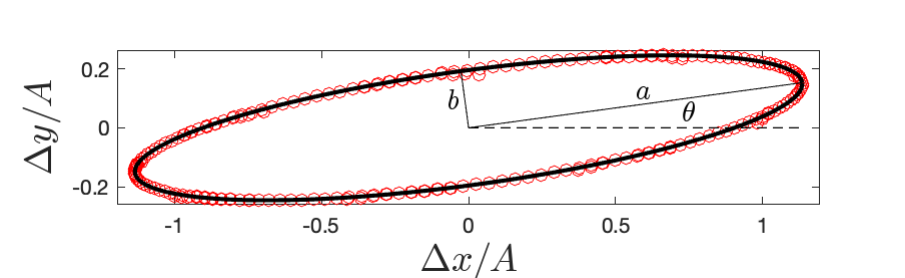} \\
    (b) \\
    \includegraphics[width=\columnwidth]{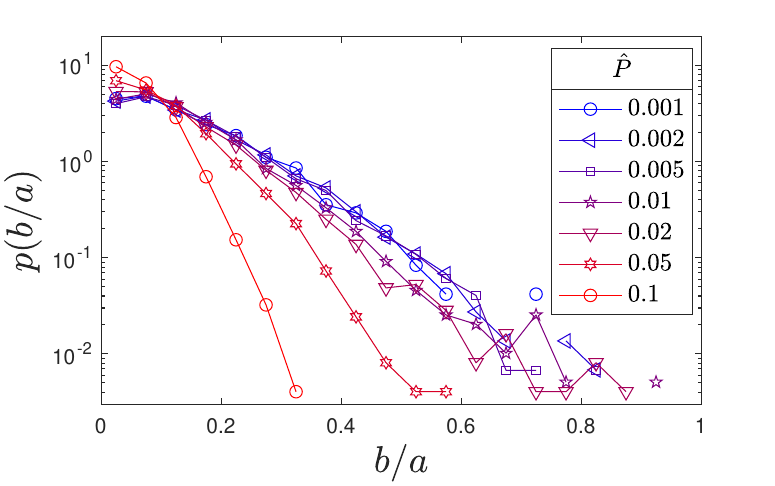} \\
    (c) \\
    \includegraphics[width=\columnwidth]{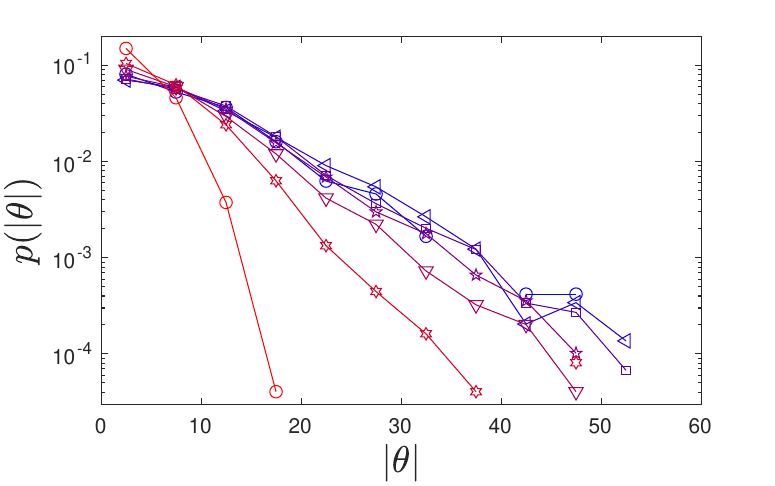}
    \caption{(a) A scatter plot of a single particle's trajectory during a simulation (red circle), along with a fitted ellipse (black line) with semi-major axis $a$, semi-minor axis $b$, and rotation angle $\theta$. This particle is typical of all particles, which are well-described by elliptical trajectories. (b,c) Probability distributions $p(b/a)$ and $p(\vert \theta \vert)$ are shown for varying $\hat{P}$ but all other parameters fixed, showing a trend toward larger aspect ratio and larger rotation angle as pressure is decreased, with asymptotic behavior for small $\hat{P}$.}
    \label{fig:2D-ellipse-stats}
\end{figure}

These emergent scaling forms likely arise from non-affine motion of grains during the local deformations associated with the traveling waves. In this context, ``non-affine'' refers to motion of individual grains that is not perfectly aligned with large-scale, average motion~\cite{utter2008experimental}. We support this claim with two figures. First, in Fig.~\ref{fig:2D-ellipse-stats}, we show that grains undergo elliptical motion with semi-major and semi-minor axes $a$ and $b$. This is in contrast to linear (affine) motion, which would be expected from a continuum picture with compressive oscillation in the $x$ direction. Additionally, we find that these ellipses have increased aspect ratio $b/a$ and increased rotation angle $\theta$ as pressure is decreased, with asymptotic behavior at low pressures (similar to the results in Figs.~\ref{fig:2D-results} and \ref{fig:2D-results-shear}).
Figure~\ref{fig:2D-ellipse-stats}(a) shows a plot of the deviation of the particles from their initial position for one grain during one simulation; the data points (red circles) trace out an elliptical shape to excellent approximation. The black line is a best-fit ellipse, with semi-major axis $a$, semi-minor axis $b$, and rotation angle $\theta$, defined as the angle between the $x$ direction and the direction of the closest semi-major axis. This particular plot is typical of all grains, i.e., all grains trace out shapes that can be fit to an ellipse with excellent fidelity.

Figure~\ref{fig:2D-ellipse-stats}(b) and (c) show probability distributions $p$ of the aspect ratio $b/a$ and the absolute value of the rotation angle $\vert \theta \vert$, since the statistics of $\theta$ are even about $\theta = 0$. For $\hat{P} = 0.1$, both distributions decay rapidly away from 0, suggesting again that at large $\hat{P}$, the motion of grains is more similar to the continuum picture (nearly linear along the direction of the applied oscillation at the boundary). However, as $\hat{P}$ is decreased, the tails of the distribution, which appear quasi-exponential, decay much more slowly, meaning there are many more particles tracing out elliptical trajectories with larger values of $b/a$ and $\vert \theta \vert$. For $\hat{P}\leq 0.005$, the probability distributions become independent of $\hat{P}$, similar to the dispersion and attenuation curves shown previously. Thus, we conclude that the anomalous scaling shown in Figs.~\ref{fig:2D-results} and \ref{fig:2D-results-shear} is directly associated with a transition from predominantly linear motion to elliptical motion.

\begin{figure}
    \raggedright
    \includegraphics[width=\columnwidth]{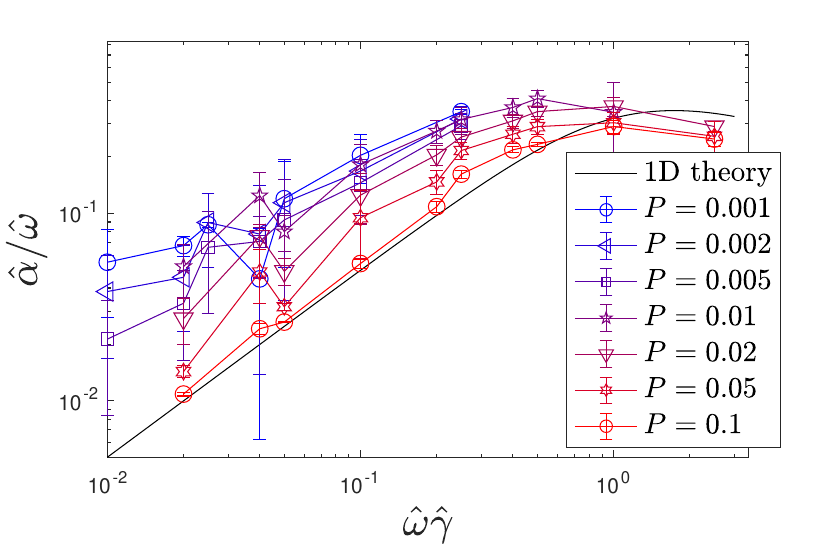}
    \caption{Attenuation in 2D with varied pressure for compressional waves, but where the tangential motion of grains is not damped. Curves are shifted but remain parallel to the 1D continuum theory (solid black line, from Eqs.~\eqref{eqn:nondim-absorption} and \eqref{eqn:nondim-dispersion}).}
    \label{fig:2D-alpha-normdash}
\end{figure}

The second figure to support our claim is Fig.~\ref{fig:2D-alpha-normdash}. Here we show dispersion and attenuation as a function of the product of dimensionless damping and frequency for simulations where we damp only the component of relative velocity that is along the grain-grain contact, i.e., we do not damp the tangential motion. This is equivalent to using $\gamma_n = \gamma$ and $\gamma_t = 0$, as discussed in Sec.~\ref{sec:sim-methods}. In such a system, we observe that the attenuation coefficient mimics the 1D theory, albeit shifted and scaled. Thus, the anomalous scaling arises when two conditions are met: first, particle oscillations have significant non-affine components that are not along the imposed oscillation direction, and, second, these non-affine motions are damped.

\subsection{3D Simulations}

\begin{figure}
    \raggedright
    \includegraphics[width=\columnwidth]{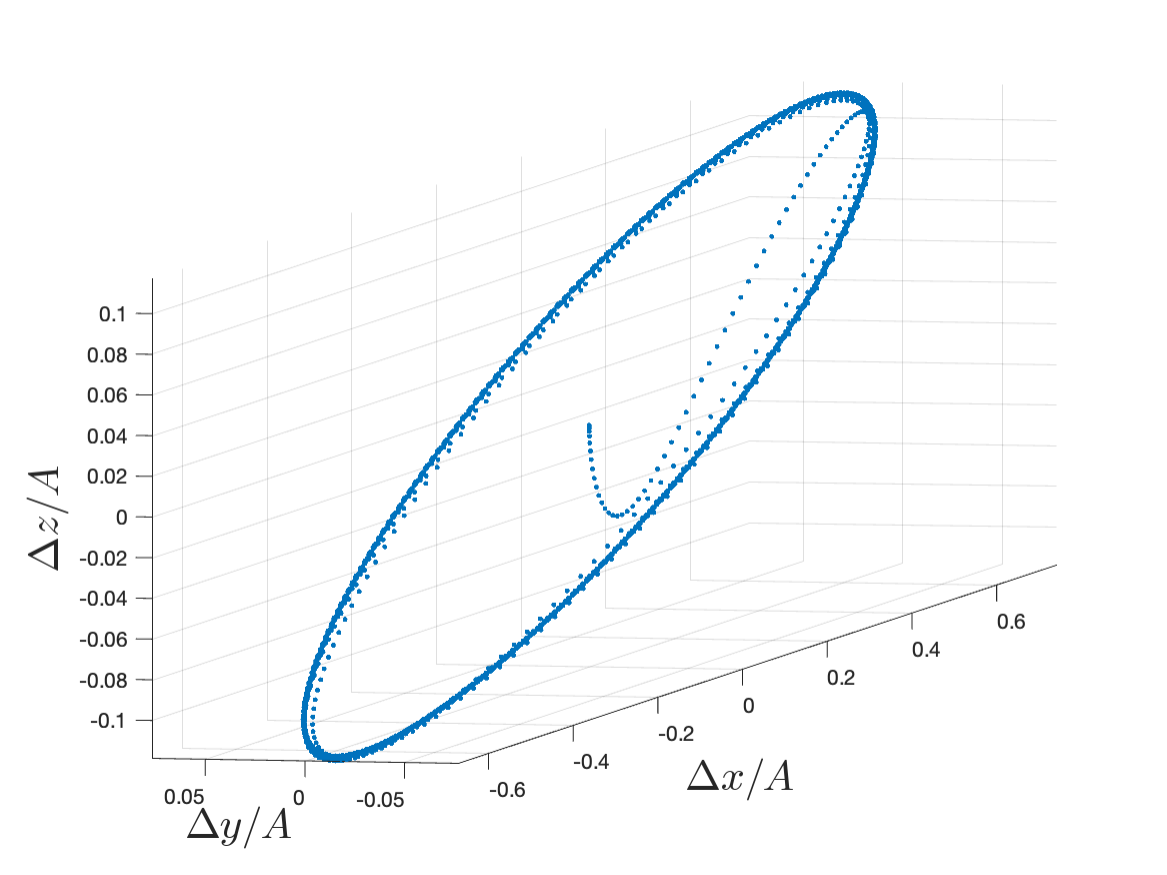}
    \caption{A scatter plot of a single particle's trajectory during a 3D simulation. This particle is typical of all particles, which are well-described by elliptical trajectories.}
    \label{fig:3D-ellipse}
\end{figure}

We repeat our 2D simulations for compressional waves in 3D. We find similar results for the packing statistics as those shown in Fig.~\ref{fig:packing-stats-2D}. We again find $Z-Z_c \propto \hat{P}^{1/2}$ but with $Z_c = 6$ for frictionless 3D spheres, as expected~\cite{van2009jamming}. We again find that particles trace out elliptical paths, as shown in Fig.~\ref{fig:3D-ellipse}. We again find that, as $\hat{P}$ is decreased, the ellipses become less aligned with the imposed oscillation at the boundary and they develop larger aspect ratios, although the summary statistics are not shown here. We save a complete analysis of particle motion, including spatial correlations, in 3D for future work.

\begin{figure}
    \raggedright
    (a) \\
    \includegraphics[width=\columnwidth]{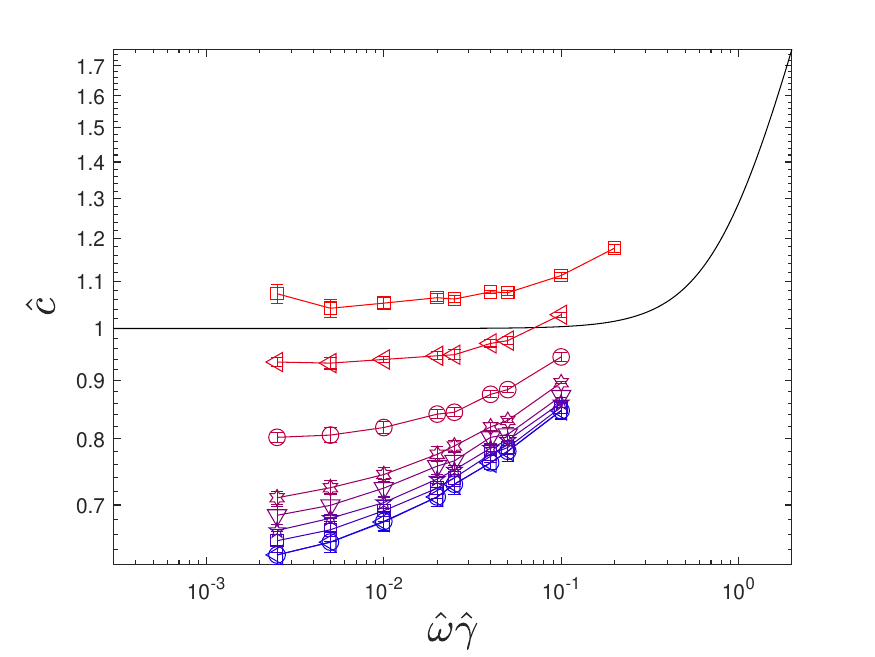}\\
    (b) \\
    \includegraphics[width=\columnwidth]{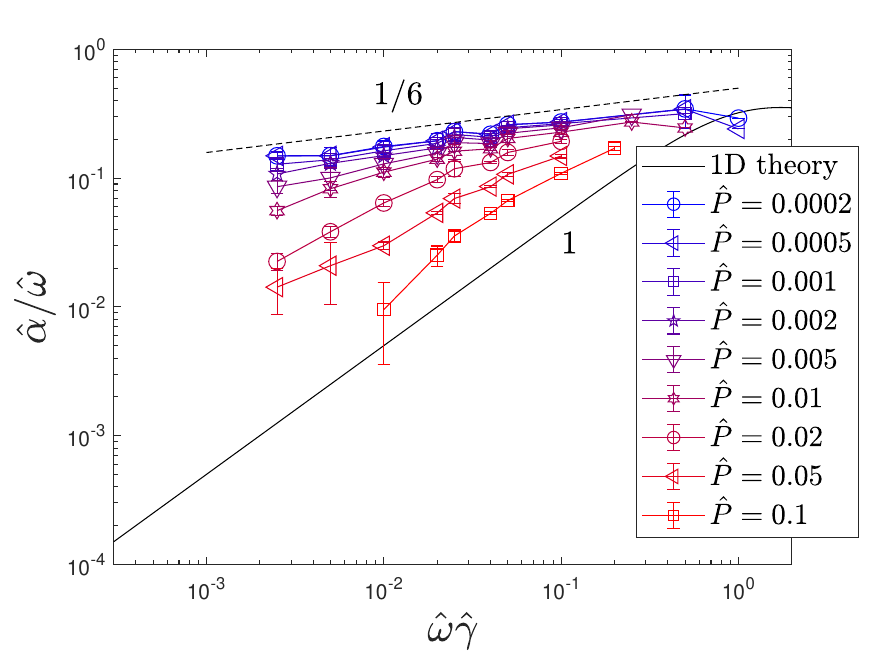}
    \caption{Dispersion and attenuation in 3D with varied pressure for compression waves. Plots show $\hat{c}$ and $\hat{\alpha}/\hat{\omega}$ versus $\hat{\omega}\hat{\gamma}$.  Solid black lines show the theoretical result from Eqs.~\eqref{eqn:nondim-absorption} and \eqref{eqn:nondim-dispersion}.}
    \label{fig:3D-results}
\end{figure}

We again consider the motion of each grain along the imposed oscillation direction and fit it to a sine wave, as in Fig.~\ref{fig:2D-single-grain-dyn}. We measure $k$ and $\alpha$ in the same way, and again plot $\hat{c}$ and $\hat{\alpha}/\hat{\omega}$ as a function of $\hat{\omega}\hat{\gamma}$ for a range of parameter values, as shown in Fig.~\ref{fig:3D-results}. We again find that $\hat{\omega}\hat{\gamma}$ is a good predictor of these two quantities, irrespective of the values of the individual parameters used to construct this combination, at least for an individual value of $\hat{P}$.

The results shown in Fig.~\ref{fig:3D-results} are very similar to the 2D case, except for the fact that $\hat{\alpha}/\hat{\omega} \propto (\hat{\omega}\hat{\gamma})^\beta$ with $\beta \approx 1/6$ in 3D, as opposed to $\beta \approx 1/3$ in 2D. This implies $\alpha \propto f^{7/6}$, which is very similar to $\alpha \propto f$. 

\section{Discussion}
\label{sec:discussion}
In this manuscript we have shown that the packing structure of the granular material may play an important role in controlling the acoustic properties of marine sediments. By studying frictionless granular packings using only linearized forces (springs and dashpots), we observe dispersion and attenuation curves that do not agree with a naive continuum approach, which predicts $\alpha \propto f^2$ for low to moderate frequencies using linear forces. Instead, we observe that the granular packing structure itself leads to emergent scaling forms with $\alpha \propto f^\beta$, where $\beta \approx 4/3$ for packings of 2D frictionless disks and $\beta \approx 7/6$ for packings of 3D frictionless spheres. The observed scaling laws are outputs of the DEM simulations and are not currently explained by any theory.

Thus, the key message of our paper is to challenge a common assumption in prior theoretical approaches, that linearized forces necessarily lead to dispersion and attenuation scaling that agrees with classic viscosity and Biot-Stoll theory, notably $\alpha \propto f^{2}$ at low $f$. Finding a minimal theoretical framework that would give rise to other scaling laws, e.g., $\alpha \propto f$, was a major factor in motivating the GS and VGS models~\cite{Buckingham1997,buckingham_wave_2000}. In contrast, our results show that linear forces can give rise to alternative scaling laws that come directly from the granular packing structure. We observe a transition from $\alpha \propto f^2$ for unphysical, high-pressure packings (with large overlaps between particles) to $\alpha \propto \beta$ with $\beta \approx 4/3$ for P-waves in 2D, $\beta \approx 1$ for S-waves in 2D, and $\beta \approx 7/6$ for P-waves in 3D. This transition occurs as the packings asymptotically approach the (more realistic) stiff-particle limit. This transition is also associated with changes in the force statistics of the packing structure (Fig.~\ref{fig:packing-stats-2D}) as well as a transition from linear to ellipitcal motion of individual grains (Fig.~\ref{fig:2D-ellipse-stats}). 

Our intent is not necessarily to advocate for any particular set of experimental data. However, we note that the most direct measurements of acoustic properties of an idealized granular packing immersed in a viscous fluid are the laboratory experiments by Hefner and Williams~\cite{hefner2006sound}. The fact that our simulation results are similar to these measurements is very encouraging. These and other results were shown above in Fig.~\ref{fig:lin-exp-data}.

There are a large number of questions that arise from our work. One line of inquiry might explore how our results would change using explicitly nonlinear grain-grain forces, like Hertzian contact mechanics or the history-dependent dashpot terms from the GS or VGS models. Additionally, more realistic granular packings, like those arising from frictional grain-grain interactions or irregular grain shapes, might give rise to changes in the scaling laws we observe. Grain-grain friction, polydispersity (i.e., a larger range of grain sizes), and irregular grain shape are known to affect the properties of the intergrain contact network~\cite{silbert2010jamming,nguyen2014effect,nguyen2015effects}. 

Future work may also try to further characterize the non-affine grain motions that are associated with the anomalous scaling behavior. For example, we have shown distributions of the ellipses traced out by grains in 2D, but have not considered whether there are spatial correlations in the orientation and aspect ratio of these ellipses. This may provide additional insight, since it is not the elliptical motion directly that increases dissipation but the relative motion between each grain and its neighbors. 

Finally, the effect of a saturating fluid beyond its contribution to the dashpot magnitude was not included in our work. Previous work\cite{Williams2009b} has shown that inertial effects of the fluid and grains must be treated properly. Including the inertia, as well as compressional and viscous effects of the saturating fluid is a key future step in relating the work presented here to marine sediment dispersion.
\section*{Conflict of Interest}
The authors declare no conflicts of interest.

\section*{Data Availability}
The data that support the findings of this study are available from the corresponding author upon reasonable request.
\acknowledgments
The authors thank Dr. Charles Holland for discussions regarding measurements of sediment dispersion. Funding was proved by the U.S. Office of Naval Research, for which the authors are grateful. Any opinions, findings, and conclusions or recommendations expressed in this publication are those of the author and do not necessarily reflect the views of the U.S. Navy.


\begin{thebibliography}{}
\def\enquote#1,{``#1,''}
\def\enxquote#1{``#1''}
\expandafter\ifx\csname url\endcsname\relax
  \def\url#1{\texttt{#1}}\fi
\expandafter\ifx\csname urlprefix\endcsname\relax\def\urlprefix{URL }\fi
\providecommand{\bibinfo}[2]{#2}
\def\plainquote#1{``#1''}
\providecommand{\noopsort}[1]{}
\providecommand{\switchargs}[2]{#2#1}
\providecommand{\dourl}[1]{\href{http://#1}{\nolinkurl{#1}}}
  \def\eatspace #1{#1}

\end{thebibliography}


\begin{thebibliography}{10}
\def\enquote#1,{``#1,''}
\def\enxquote#1{``#1''}
\expandafter\ifx\csname url\endcsname\relax
  \def\url#1{\texttt{#1}}\fi
\expandafter\ifx\csname urlprefix\endcsname\relax\def\urlprefix{URL }\fi
\providecommand{\bibinfo}[2]{#2}
\def\plainquote#1{``#1''}
\providecommand{\noopsort}[1]{}
\providecommand{\switchargs}[2]{#2#1}
\providecommand{\dourl}[1]{\href{http://#1}{\nolinkurl{#1}}}
  \def\eatspace #1{#1}

\bibitem{ballard2017acoustics}
\bibinfo{author}{M.~S. Ballard} and \bibinfo{author}{K.~Lee},
  \enquote{\bibinfo{title}{The acoustics of marine sediments}},
  \bibinfo{journal}{Acoust. Today} \textbf{13}(3), \bibinfo{pages}{11--18}
  (\bibinfo{year}{2017}).

\bibitem{daraio2005strongly}
\bibinfo{author}{C.~Daraio}, \bibinfo{author}{V.~Nesterenko},
  \bibinfo{author}{E.~Herbold}, and \bibinfo{author}{S.~Jin},
  \enquote{\bibinfo{title}{Strongly nonlinear waves in a chain of teflon
  beads}},  \bibinfo{journal}{Physical Review E} \textbf{72}(1),
  \bibinfo{pages}{016603} (\bibinfo{year}{2005}).

\bibitem{gomez2012shocks}
\bibinfo{author}{L.~R. G{\'o}mez}, \bibinfo{author}{A.~M. Turner},
  \bibinfo{author}{M.~van Hecke}, and \bibinfo{author}{V.~Vitelli},
  \enquote{\bibinfo{title}{Shocks near jamming}},  \bibinfo{journal}{Physical
  review letters} \textbf{108}(5), \bibinfo{pages}{058001}
  (\bibinfo{year}{2012}).

\bibitem{force_schemes}
\bibinfo{author}{J.~Shäfer}, \bibinfo{author}{S.~Dippel}, and
  \bibinfo{author}{D.~Wolf}, \enquote{\bibinfo{title}{Force schemes in
  simulations of granular materials}},  \bibinfo{journal}{Journal de Physique
  I} \textbf{6} (\bibinfo{year}{1996}) \dodoi{10.1051/jp1:1996129}.

\bibitem{majmudar2005contact}
\bibinfo{author}{T.~S. Majmudar} and \bibinfo{author}{R.~P. Behringer},
  \enquote{\bibinfo{title}{Contact force measurements and stress-induced
  anisotropy in granular materials}},  \bibinfo{journal}{nature}
  \textbf{435}(7045), \bibinfo{pages}{1079--1082} (\bibinfo{year}{2005}).

\bibitem{Silbert_PRL_2005}
\bibinfo{author}{L.~E. Silbert}, \bibinfo{author}{A.~J. Liu}, and
  \bibinfo{author}{S.~R. Nagel}, \enquote{\bibinfo{title}{Vibrations and
  diverging length scales near the unjamming transition}},
  \bibinfo{journal}{Phys. Rev. Lett.} \textbf{95}, \bibinfo{pages}{098301}
  (\bibinfo{year}{2005})
  \dourl{https://link.aps.org/doi/10.1103/PhysRevLett.95.098301}
  \dodoi{10.1103/PhysRevLett.95.098301}.

\bibitem{chotiros2017Acoustics}
\bibinfo{author}{N.~P. Chotiros}, \emph{\bibinfo{title}{Acoustics of the Seabed
  as a Poroelastic Medium}}  (\bibinfo{publisher}{Springer},
  \bibinfo{year}{2017}).

\bibitem{Williams2009b}
\bibinfo{author}{K.~L. Williams}, \enquote{\bibinfo{title}{Sand acoustics: The
  effective density fluid model, pierce/carey expressions, and inferences for
  porous media modeling}},  \bibinfo{journal}{The Journal of the Acoustical
  Society of America} \textbf{125}(4), \bibinfo{pages}{EL164--EL170}
  (\bibinfo{year}{2009}) \dodoi{10.1121/1.3097681}.

\bibitem{Holland2012}
\bibinfo{author}{C.~W. Holland}, \bibinfo{author}{P.~L. Nielsen},
  \bibinfo{author}{J.~Dettmer}, and \bibinfo{author}{S.~Dosso},
  \enquote{\bibinfo{title}{Resolving meso-scale seabed variability using
  reflection measurements from an autonomous underwater vehicle}},
  \bibinfo{journal}{The Journal of the Acoustical Society of America}
  \textbf{131}(2), \bibinfo{pages}{1066--1078} (\bibinfo{year}{2012})
  \dodoi{10.1121/1.3672696}.

\bibitem{Dettmer2010}
\bibinfo{author}{J.~Dettmer}, \bibinfo{author}{S.~E. Dosso}, and
  \bibinfo{author}{C.~W. Holland}, \enquote{\bibinfo{title}{Trans-dimensional
  geoacoustic inversion}},  \bibinfo{journal}{The Journal of the Acoustical
  Society of America} \textbf{128}(6), \bibinfo{pages}{3393--3405}
  (\bibinfo{year}{2010}) \dodoi{10.1121/1.3500674}.

\bibitem{Quijano2012}
\bibinfo{author}{J.~E. Quijano}, \bibinfo{author}{S.~E. Dosso},
  \bibinfo{author}{J.~Dettmer}, \bibinfo{author}{L.~M. Zurk},
  \bibinfo{author}{M.~Siderius}, and \bibinfo{author}{C.~H. Harrison},
  \enquote{\bibinfo{title}{Bayesian geoacoustic inversion using wind-driven
  ambient noise}},  \bibinfo{journal}{The Journal of the Acoustical Society of
  America} \textbf{131}(4), \bibinfo{pages}{2658--2667} (\bibinfo{year}{2012})
  \dodoi{10.1121/1.3688482}.

\bibitem{Holland2017}
\bibinfo{author}{C.~W. Holland}, \bibinfo{author}{S.~Pinson},
  \bibinfo{author}{C.~M. Smith}, \bibinfo{author}{P.~C. Hines},
  \bibinfo{author}{D.~R. Olson}, \bibinfo{author}{S.~E. Dosso}, and
  \bibinfo{author}{J.~Dettmer}, \enquote{\bibinfo{title}{Seabed structure
  inferences from {TREX}13 reflection measurements}},  \bibinfo{journal}{{IEEE}
  Journal of Oceanic Engineering} \textbf{42}(2), \bibinfo{pages}{268--288}
  (\bibinfo{year}{2017}) \dodoi{10.1109/joe.2017.2658418}.

\bibitem{Olson2023}
\bibinfo{author}{D.~R. Olson}, \enquote{\bibinfo{title}{The effect of seafloor
  roughness on passive estimates of the seabed reflection coefficient}},
  \bibinfo{journal}{The Journal of the Acoustical Society of America}
  \textbf{153}(1), \bibinfo{pages}{586--601} (\bibinfo{year}{2023})
  \dodoi{10.1121/10.0016846}.

\bibitem{Bonnel2013}
\bibinfo{author}{J.~Bonnel}, \bibinfo{author}{S.~E. Dosso}, and
  \bibinfo{author}{N.~Ross~Chapman}, \enquote{\bibinfo{title}{Bayesian
  geoacoustic inversion of single hydrophone light bulb data using warping
  dispersion analysis}},  \bibinfo{journal}{The Journal of the Acoustical
  Society of America} \textbf{134}(1), \bibinfo{pages}{120--130}
  (\bibinfo{year}{2013}) \dodoi{10.1121/1.4809678}.

\bibitem{tan_ambient_2022}
\bibinfo{author}{T.~W. Tan} and \bibinfo{author}{O.~A. Godin},
  \enquote{\bibinfo{title}{Ambient sound directionality and rapid estimation of
  empirical {Green}’s functions in a coastal ocean}},  \bibinfo{journal}{The
  Journal of the Acoustical Society of America} \textbf{152}(4),
  \bibinfo{pages}{A153} (\bibinfo{year}{2022})
  \dourl{https://doi.org/10.1121/10.0015861} \dodoi{10.1121/10.0015861}.

\bibitem{Holland2013}
\bibinfo{author}{C.~W. Holland} and \bibinfo{author}{J.~Dettmer},
  \enquote{\bibinfo{title}{In situ sediment dispersion estimates in the
  presence of discrete layers and gradients}},  \bibinfo{journal}{The Journal
  of the Acoustical Society of America} \textbf{133}(1),
  \bibinfo{pages}{50--61} (\bibinfo{year}{2013}) \dodoi{10.1121/1.4765300}.

\bibitem{Porter1985}
\bibinfo{author}{M.~B. Porter} and \bibinfo{author}{E.~L. Reiss},
  \enquote{\bibinfo{title}{A numerical method for bottom interacting ocean
  acoustic normal modes}},  \bibinfo{journal}{The Journal of the Acoustical
  Society of America} \textbf{77}(5), \bibinfo{pages}{1760--1767}
  (\bibinfo{year}{1985}) \dodoi{10.1121/1.391925}.

\bibitem{Holland1998}
\bibinfo{author}{C.~W. Holland} and \bibinfo{author}{P.~Neumann},
  \enquote{\bibinfo{title}{Sub-bottom scattering: A modeling approach}},
  \bibinfo{journal}{J. Acoust. Soc. America} \textbf{104}(3),
  \bibinfo{pages}{1363--1373} (\bibinfo{year}{1998}) \dodoi{10.1121/1.424345}.

\bibitem{LePage2000}
\bibinfo{author}{K.~D. LePage} and \bibinfo{author}{H.~Schmidt},
  \enquote{\bibinfo{title}{Spectral integral representations of volume
  scattering in sediments in layered waveguides}},  \bibinfo{journal}{The
  Journal of the Acoustical Society of America} \textbf{108}(4),
  \bibinfo{pages}{1557--1567} (\bibinfo{year}{2000}) \dodoi{10.1121/1.1289370}.

\bibitem{LePage2003}
\bibinfo{author}{K.~D. LePage} and \bibinfo{author}{H.~Schmidt},
  \enquote{\bibinfo{title}{Spectral integral representations of monostatic
  backscattering from three-dimensional distributions of sediment volume
  inhomogeneities}},  \bibinfo{journal}{J. Acoust. Soc. America}
  \textbf{113}(2), \bibinfo{pages}{789--799} (\bibinfo{year}{2003})
  \dodoi{10.1121/1.1528625}.

\bibitem{Tang2017}
\bibinfo{author}{D.~Tang} and \bibinfo{author}{D.~Jackson},
  \enquote{\bibinfo{title}{A time-domain model for seafloor scattering}},
  \bibinfo{journal}{The Journal of the Acoustical Society of America}
  \textbf{142}(5), \bibinfo{pages}{2968--2978} (\bibinfo{year}{2017})
  \dodoi{10.1121/1.5009932}.

\bibitem{Olson2020}
\bibinfo{author}{D.~R. Olson} and \bibinfo{author}{C.~W. Holland},
  \enquote{\bibinfo{title}{Fast computation of time-domain scattering by an
  inhomogeneous stratified seafloor}},  \bibinfo{journal}{The Journal of the
  Acoustical Society of America} \textbf{147}(1), \bibinfo{pages}{191--204}
  (\bibinfo{year}{2020}) \dodoi{10.1121/10.0000570}.

\bibitem{Jackson2020}
\bibinfo{author}{D.~Jackson} and \bibinfo{author}{D.~R. Olson},
  \enquote{\bibinfo{title}{The small-slope approximation for layered, fluid
  seafloors}},  \bibinfo{journal}{The Journal of the Acoustical Society of
  America} \textbf{147}(1), \bibinfo{pages}{56--73} (\bibinfo{year}{2020})
  \dourl{https://doi.org/10.1121/10.0000470} \dodoi{10.1121/10.0000470}.

\bibitem{Olson2020a}
\bibinfo{author}{D.~R. Olson} and \bibinfo{author}{D.~Jackson},
  \enquote{\bibinfo{title}{Scattering from layered seafloors: Comparisons
  between theory and integral equations}},  \bibinfo{journal}{The Journal of
  the Acoustical Society of America} \textbf{148}(4),
  \bibinfo{pages}{2086--2095} (\bibinfo{year}{2020})
  \dodoi{10.1121/10.0002164}.

\bibitem{Abraham2019}
\bibinfo{author}{D.~A. Abraham}, \emph{\bibinfo{title}{Underwater Acoustic
  Signal Processing}}  (\bibinfo{publisher}{Springer International Publishing},
  \bibinfo{year}{2019}).

\bibitem{zhou_low-frequency_2009}
\bibinfo{author}{J.-X. Zhou}, \bibinfo{author}{X.-Z. Zhang}, and
  \bibinfo{author}{D.~P. Knobles}, \enquote{\bibinfo{title}{Low-frequency
  geoacoustic model for the effective properties of sandy seabottoms}},
  \bibinfo{journal}{The Journal of the Acoustical Society of America}
  \textbf{125}(5), \bibinfo{pages}{2847--2866} (\bibinfo{year}{2009})
  \dourl{https://doi.org/10.1121/1.3089218} \dodoi{10.1121/1.3089218}
  \bibinfo{note}{publisher: Acoustical Society of America}.

\bibitem{biot_theory_2005-1}
\bibinfo{author}{M.~Biot}, \enquote{\bibinfo{title}{Theory of {propagation} of
  {elastic} {waves} in a {fluid-saturated} {porous} {solid}. {I}.
  {Low-frequency} {range}}},  \bibinfo{journal}{The Journal of the Acoustical
  Society of America}  (\bibinfo{year}{1956})
  \dourl{https://pubs.aip.org/asa/jasa/article/28/2/168/737238/Theory-of-Propagation-of-Elastic-Waves-in-a-Fluid}.

\bibitem{biot_theory_2005}
\bibinfo{author}{M.~A. Biot}, \enquote{\bibinfo{title}{Theory of {propagation}
  of {elastic} {waves} in a {fluid-saturated} {porous} {solid}. {II}. {Higher}
  {frequency} {range}}},  \bibinfo{journal}{The Journal of the Acoustical
  Society of America} \textbf{28}(2), \bibinfo{pages}{179--191}
  (\bibinfo{year}{1955}) \dourl{https://doi.org/10.1121/1.1908241}
  \dodoi{10.1121/1.1908241}.

\bibitem{biot_generalized_2005}
\bibinfo{author}{M.~A. Biot}, \enquote{\bibinfo{title}{Generalized {theory} of
  {acoustic} {propagation} in {porous} {dissipative} {media}}},
  \bibinfo{journal}{The Journal of the Acoustical Society of America}
  \textbf{34}(9A), \bibinfo{pages}{1254--1264} (\bibinfo{year}{1962})
  \dourl{https://doi.org/10.1121/1.1918315} \dodoi{10.1121/1.1918315}.

\bibitem{stoll_acoustic_1977}
\bibinfo{author}{R.~D. Stoll}, \enquote{\bibinfo{title}{Acoustic waves in ocean
  sediments}},  \bibinfo{journal}{GEOPHYSICS} \textbf{42}(4),
  \bibinfo{pages}{715--725} (\bibinfo{year}{1977})
  \dourl{https://library.seg.org/doi/10.1190/1.1440741}
  \dodoi{10.1190/1.1440741} \bibinfo{note}{publisher: Society of Exploration
  Geophysicists}.

\bibitem{stoll_theoretical_1980}
\bibinfo{author}{R.~D. Stoll}, \enquote{\bibinfo{title}{Theoretical aspects of
  sound transmission in sediments}},  \bibinfo{journal}{The Journal of the
  Acoustical Society of America} \textbf{68}(5), \bibinfo{pages}{1341--1350}
  (\bibinfo{year}{1980}) \dourl{https://doi.org/10.1121/1.385101}
  \dodoi{10.1121/1.385101}.

\bibitem{stoll_reflection_1981}
\bibinfo{author}{R.~D. Stoll} and \bibinfo{author}{T.~K. Kan},
  \enquote{\bibinfo{title}{Reflection of acoustic waves at a water–sediment
  interface}},  \bibinfo{journal}{The Journal of the Acoustical Society of
  America} \textbf{70}(1), \bibinfo{pages}{149--156} (\bibinfo{year}{1981})
  \dourl{https://doi.org/10.1121/1.386692} \dodoi{10.1121/1.386692}.

\bibitem{stoll_marine_1985}
\bibinfo{author}{R.~D. Stoll}, \enquote{\bibinfo{title}{Marine sediment
  acoustics}},  \bibinfo{journal}{The Journal of the Acoustical Society of
  America} \textbf{77}(5), \bibinfo{pages}{1789--1799} (\bibinfo{year}{1985})
  \dourl{https://doi.org/10.1121/1.391928} \dodoi{10.1121/1.391928}.

\bibitem{Godin_2021}
\bibinfo{author}{O.~A. Godin}, \enquote{\bibinfo{title}{Shear waves and sound
  attenuation in underwater waveguides}},  \bibinfo{journal}{The Journal of the
  Acoustical Society of America} \textbf{149}(5), \bibinfo{pages}{3586--3598}
  (\bibinfo{year}{2021}) \dodoi{10.1121/10.0004999}.

\bibitem{godin_effects_2023}
\bibinfo{author}{O.~A. Godin}, \enquote{\bibinfo{title}{Effects of weak shear
  rigidity of the seabed on sound propagation in a range-dependent ocean}},
  \bibinfo{journal}{The Journal of the Acoustical Society of America}
  \textbf{153}(3\_supplement), \bibinfo{pages}{A86} (\bibinfo{year}{2023})
  \dourl{https://doi.org/10.1121/10.0018259} \dodoi{10.1121/10.0018259}.

\bibitem{Hamilton1980}
\bibinfo{author}{E.~L. Hamilton}, \enquote{\bibinfo{title}{Geoacoustic modeling
  of the sea floor}},  \bibinfo{journal}{J. Acoust. Soc. America}
  \textbf{68}(5), \bibinfo{pages}{1313--1340} (\bibinfo{year}{1980})
  \dodoi{10.1121/1.385100}.

\bibitem{williams2002comparison}
\bibinfo{author}{K.~L. Williams}, \bibinfo{author}{D.~R. Jackson},
  \bibinfo{author}{E.~I. Thorsos}, \bibinfo{author}{D.~Tang}, and
  \bibinfo{author}{S.~G. Schock}, \enquote{\bibinfo{title}{Comparison of sound
  speed and attenuation measured in a sandy sediment to predictions based on
  the biot theory of porous media}},  \bibinfo{journal}{IEEE journal of oceanic
  engineering} \textbf{27}(3), \bibinfo{pages}{413--428}
  (\bibinfo{year}{2002}).

\bibitem{Hefner2009}
\bibinfo{author}{B.~Hefner}, \bibinfo{author}{D.~Jackson},
  \bibinfo{author}{K.~Williams}, and \bibinfo{author}{E.~Thorsos},
  \enquote{\bibinfo{title}{Mid- to high-frequency acoustic penetration and
  propagation measurements in a sandy sediment}},  \bibinfo{journal}{IEEE
  Journal of Oceanic Engineering} \textbf{34}(4), \bibinfo{pages}{372--387}
  (\bibinfo{year}{2009}) \dodoi{10.1109/joe.2009.2028410}.

\bibitem{Turgut1990}
\bibinfo{author}{A.~Turgut} and \bibinfo{author}{T.~Yamamoto},
  \enquote{\bibinfo{title}{Measurements of acoustic wave velocities and
  attenuation in marine sediments}},  \bibinfo{journal}{The Journal of the
  Acoustical Society of America} \textbf{87}(6), \bibinfo{pages}{2376--2383}
  (\bibinfo{year}{1990}) \dodoi{10.1121/1.399084}.

\bibitem{Simpson2003}
\bibinfo{author}{H.~J. Simpson}, \bibinfo{author}{B.~H. Houston},
  \bibinfo{author}{S.~W. Liskey}, \bibinfo{author}{P.~A. Frank},
  \bibinfo{author}{A.~R. Berdoz}, \bibinfo{author}{L.~A. Kraus},
  \bibinfo{author}{C.~K. Frederickson}, and \bibinfo{author}{S.~Stanic},
  \enquote{\bibinfo{title}{At-sea measurements of sound penetration into
  sediments using a buried vertical synthetic array}},  \bibinfo{journal}{The
  Journal of the Acoustical Society of America} \textbf{114}(3),
  \bibinfo{pages}{1281--1290} (\bibinfo{year}{2003}) \dodoi{10.1121/1.1594192}.

\bibitem{Wingham1985}
\bibinfo{author}{D.~J. Wingham}, \enquote{\bibinfo{title}{The dispersion of
  sound in sediment}},  \bibinfo{journal}{The Journal of the Acoustical Society
  of America} \textbf{78}(5), \bibinfo{pages}{1757--1760}
  (\bibinfo{year}{1985}) \dodoi{10.1121/1.392761}.

\bibitem{hefner2006sound}
\bibinfo{author}{B.~T. Hefner} and \bibinfo{author}{K.~L. Williams},
  \enquote{\bibinfo{title}{Sound speed and attenuation measurements in
  unconsolidated glass-bead sediments saturated with viscous pore fluids}},
  \bibinfo{journal}{The Journal of the Acoustical Society of America}
  \textbf{120}(5), \bibinfo{pages}{2538--2549} (\bibinfo{year}{2006}).

\bibitem{Lee2007}
\bibinfo{author}{K.~I. Lee}, \bibinfo{author}{V.~F. Humphrey},
  \bibinfo{author}{B.-N. Kim}, and \bibinfo{author}{S.~W. Yoon},
  \enquote{\bibinfo{title}{Frequency dependencies of phase velocity and
  attenuation coefficient in a water-saturated sandy sediment from
  0.3to1.0mhz}},  \bibinfo{journal}{The Journal of the Acoustical Society of
  America} \textbf{121}(5), \bibinfo{pages}{2553--2558} (\bibinfo{year}{2007})
  \dodoi{10.1121/1.2713690}.

\bibitem{Sessarego2008}
\bibinfo{author}{J.-P. Sessarego}, \bibinfo{author}{A.~Ivakin}, and
  \bibinfo{author}{D.~Ferrand}, \enquote{\bibinfo{title}{Frequency dependence
  of phase speed, group speed, and attenuation in water-saturated sand:
  Laboratory experiments}},  \bibinfo{journal}{IEEE Journal of Oceanic
  Engineering} \textbf{33}(4), \bibinfo{pages}{359--366} (\bibinfo{year}{2008})
  \dodoi{10.1109/joe.2008.927584}.

\bibitem{Hare2020}
\bibinfo{author}{J.~Hare} and \bibinfo{author}{A.~E. Hay},
  \enquote{\bibinfo{title}{Phase speed in water-saturated sand and glass beads
  at {MHz} frequencies}},  \bibinfo{journal}{The Journal of the Acoustical
  Society of America} \textbf{148}(4), \bibinfo{pages}{2301--2310}
  (\bibinfo{year}{2020}) \dodoi{10.1121/10.0002250}.

\bibitem{Buckingham1997}
\bibinfo{author}{M.~J. Buckingham}, \enquote{\bibinfo{title}{Theory of acoustic
  attenuation, dispersion, and pulse propagation in unconsolidated granular
  materials including marine sediments}},  \bibinfo{journal}{The Journal of the
  Acoustical Society of America} \textbf{102}(5), \bibinfo{pages}{2579--2596}
  (\bibinfo{year}{1997}) \dodoi{10.1121/1.420313}.

\bibitem{buckingham_wave_2000}
\bibinfo{author}{M.~J. Buckingham}, \enquote{\bibinfo{title}{Wave propagation,
  stress relaxation, and grain-to-grain shearing in saturated, unconsolidated
  marine sediments}},  \bibinfo{journal}{The Journal of the Acoustical Society
  of America} \textbf{108}(6), \bibinfo{pages}{2796--2815}
  (\bibinfo{year}{2000}) \dourl{https://doi.org/10.1121/1.1322018}
  \dodoi{10.1121/1.1322018}.

\bibitem{buckingham_pore-fluid_2007}
\bibinfo{author}{M.~J. Buckingham}, \enquote{\bibinfo{title}{On pore-fluid
  viscosity and the wave properties of saturated granular materials including
  marine sediments}},  \bibinfo{journal}{The Journal of the Acoustical Society
  of America} \textbf{108}(6) (\bibinfo{year}{2007})
  \dourl{https://pubs.aip.org/asa/jasa/article/122/3/1486/853009/On-pore-fluid-viscosity-and-the-wave-properties-of}.

\bibitem{jaeger1996granular}
\bibinfo{author}{H.~M. Jaeger}, \bibinfo{author}{S.~R. Nagel}, and
  \bibinfo{author}{R.~P. Behringer}, \enquote{\bibinfo{title}{Granular solids,
  liquids, and gases}},  \bibinfo{journal}{Reviews of modern physics}
  \textbf{68}(4), \bibinfo{pages}{1259} (\bibinfo{year}{1996}).

\bibitem{forterre2008flows}
\bibinfo{author}{Y.~Forterre} and \bibinfo{author}{O.~Pouliquen},
  \enquote{\bibinfo{title}{Flows of dense granular media}},
  \bibinfo{journal}{Annu. Rev. Fluid Mech.} \textbf{40}, \bibinfo{pages}{1--24}
  (\bibinfo{year}{2008}).

\bibitem{van2009jamming}
\bibinfo{author}{M.~van Hecke}, \enquote{\bibinfo{title}{Jamming of soft
  particles: Geometry, mechanics, scaling and isostaticity}},
  \bibinfo{journal}{Journal of Physics: Condensed Matter} \textbf{22}(3),
  \bibinfo{pages}{033101} (\bibinfo{year}{2009}).

\bibitem{liu2010jamming}
\bibinfo{author}{A.~J. Liu} and \bibinfo{author}{S.~R. Nagel},
  \enquote{\bibinfo{title}{The jamming transition and the marginally jammed
  solid}},  \bibinfo{journal}{Annu. Rev. Condens. Matter Phys.} \textbf{1}(1),
  \bibinfo{pages}{347--369} (\bibinfo{year}{2010}).

\bibitem{behringer2018physics}
\bibinfo{author}{R.~P. Behringer} and \bibinfo{author}{B.~Chakraborty},
  \enquote{\bibinfo{title}{The physics of jamming for granular materials: a
  review}},  \bibinfo{journal}{Reports on Progress in Physics} \textbf{82}(1),
  \bibinfo{pages}{012601} (\bibinfo{year}{2018}).

\bibitem{amon2017preface}
\bibinfo{author}{A.~Amon}, \bibinfo{author}{P.~Born}, \bibinfo{author}{K.~E.
  Daniels}, \bibinfo{author}{J.~A. Dijksman}, \bibinfo{author}{K.~Huang},
  \bibinfo{author}{D.~Parker}, \bibinfo{author}{M.~Schr{\"o}ter},
  \bibinfo{author}{R.~Stannarius}, and \bibinfo{author}{A.~Wierschem},
  \enquote{\bibinfo{title}{Preface: Focus on imaging methods in granular
  physics}},  \bibinfo{journal}{Review of Scientific Instruments}
  \textbf{88}(5) (\bibinfo{year}{2017}).

\bibitem{radjai2017modeling}
\bibinfo{author}{F.~Radjai}, \bibinfo{author}{J.-N. Roux}, and
  \bibinfo{author}{A.~Daouadji}, \enquote{\bibinfo{title}{Modeling granular
  materials: century-long research across scales}},  \bibinfo{journal}{Journal
  of engineering mechanics} \textbf{143}(4), \bibinfo{pages}{04017002}
  (\bibinfo{year}{2017}).

\bibitem{kamrin2024advances}
\bibinfo{author}{K.~Kamrin}, \bibinfo{author}{K.~M. Hill},
  \bibinfo{author}{D.~I. Goldman}, and \bibinfo{author}{J.~E. Andrade},
  \enquote{\bibinfo{title}{Advances in modeling dense granular media}},
  \bibinfo{journal}{Annual Review of Fluid Mechanics} \textbf{56},
  \bibinfo{pages}{215--240} (\bibinfo{year}{2024}).

\bibitem{Wyart_PRE_2005}
\bibinfo{author}{M.~Wyart}, \bibinfo{author}{L.~E. Silbert},
  \bibinfo{author}{S.~R. Nagel}, and \bibinfo{author}{T.~A. Witten},
  \enquote{\bibinfo{title}{Effects of compression on the vibrational modes of
  marginally jammed solids}},  \bibinfo{journal}{Phys. Rev. E} \textbf{72},
  \bibinfo{pages}{051306} (\bibinfo{year}{2005})
  \dourl{https://link.aps.org/doi/10.1103/PhysRevE.72.051306}
  \dodoi{10.1103/PhysRevE.72.051306}.

\bibitem{saitoh2023sound}
\bibinfo{author}{K.~Saitoh}, \bibinfo{author}{K.~Taghizadeh}, and
  \bibinfo{author}{S.~Luding}, \enquote{\bibinfo{title}{Sound characteristics
  of disordered granular disks: effects of contact damping}},
  \bibinfo{journal}{Frontiers in Physics} \textbf{11}, \bibinfo{pages}{1192270}
  (\bibinfo{year}{2023}).

\bibitem{owens_sound_2011}
\bibinfo{author}{E.~T. Owens} and \bibinfo{author}{K.~E. Daniels},
  \enquote{\bibinfo{title}{Sound propagation and force chains in granular
  materials}},  \bibinfo{journal}{Europhysics Letters} \textbf{94}(5),
  \bibinfo{pages}{54005} (\bibinfo{year}{2011})
  \dourl{https://dx.doi.org/10.1209/0295-5075/94/54005}
  \dodoi{10.1209/0295-5075/94/54005}.

\bibitem{Schreck_PRL_2011}
\bibinfo{author}{C.~F. Schreck}, \bibinfo{author}{T.~Bertrand},
  \bibinfo{author}{C.~S. O'Hern}, and \bibinfo{author}{M.~D. Shattuck},
  \enquote{\bibinfo{title}{Repulsive contact interactions make jammed
  particulate systems inherently nonharmonic}},  \bibinfo{journal}{Phys. Rev.
  Lett.} \textbf{107}, \bibinfo{pages}{078301} (\bibinfo{year}{2011})
  \dourl{https://link.aps.org/doi/10.1103/PhysRevLett.107.078301}
  \dodoi{10.1103/PhysRevLett.107.078301}.

\bibitem{silbert2010jamming}
\bibinfo{author}{L.~E. Silbert}, \enquote{\bibinfo{title}{Jamming of frictional
  spheres and random loose packing}},  \bibinfo{journal}{Soft Matter}
  \textbf{6}(13), \bibinfo{pages}{2918--2924} (\bibinfo{year}{2010}).

\bibitem{nguyen2014effect}
\bibinfo{author}{D.-H. Nguyen}, \bibinfo{author}{{\'E}.~Az{\'e}ma},
  \bibinfo{author}{F.~Radjai}, and \bibinfo{author}{P.~Sornay},
  \enquote{\bibinfo{title}{Effect of size polydispersity versus particle shape
  in dense granular media}},  \bibinfo{journal}{Physical Review E}
  \textbf{90}(1), \bibinfo{pages}{012202} (\bibinfo{year}{2014}).

\bibitem{nguyen2015effects}
\bibinfo{author}{D.-H. Nguyen}, \bibinfo{author}{E.~Az{\'e}ma},
  \bibinfo{author}{P.~Sornay}, and \bibinfo{author}{F.~Radjai},
  \enquote{\bibinfo{title}{Effects of shape and size polydispersity on strength
  properties of granular materials}},  \bibinfo{journal}{Physical Review E}
  \textbf{91}(3), \bibinfo{pages}{032203} (\bibinfo{year}{2015}).

\bibitem{coste1999validity}
\bibinfo{author}{C.~Coste} and \bibinfo{author}{B.~Gilles},
  \enquote{\bibinfo{title}{On the validity of hertz contact law for granular
  material acoustics}},  \bibinfo{journal}{The European Physical Journal
  B-Condensed Matter and Complex Systems} \textbf{7}, \bibinfo{pages}{155--168}
  (\bibinfo{year}{1999}).

\bibitem{Garrett2020a}
\bibinfo{author}{S.~L. Garrett}, \emph{\bibinfo{title}{Understanding
  Acoustics}}  (\bibinfo{publisher}{Springer International Publishing},
  \bibinfo{year}{2020}).

\bibitem{shitikova2022models}
\bibinfo{author}{M.~V. Shitikova} and \bibinfo{author}{A.~I. Krusser},
  \enquote{\bibinfo{title}{Models of viscoelastic materials: a review on
  historical development and formulation}},  \bibinfo{journal}{Theoretical
  Analyses, Computations, and Experiments of Multiscale Materials: A Tribute to
  Francesco dell’Isola} \bibinfo{pages}{285--326} (\bibinfo{year}{2022}).

\bibitem{binder2004molecular}
\bibinfo{author}{K.~Binder}, \bibinfo{author}{J.~Horbach},
  \bibinfo{author}{W.~Kob}, \bibinfo{author}{W.~Paul}, and
  \bibinfo{author}{F.~Varnik}, \enquote{\bibinfo{title}{Molecular dynamics
  simulations}},  \bibinfo{journal}{Journal of Physics: Condensed Matter}
  \textbf{16}(5), \bibinfo{pages}{S429} (\bibinfo{year}{2004}).

\bibitem{Montaine2011}
\bibinfo{author}{M.~Montaine}, \bibinfo{author}{M.~Heckel},
  \bibinfo{author}{C.~Kruelle}, \bibinfo{author}{T.~Schwager}, and
  \bibinfo{author}{T.~Pöschel}, \enquote{\bibinfo{title}{Coefficient of
  restitution as a fluctuating quantity}},  \bibinfo{journal}{Physical Review
  E} \textbf{84}(4), \bibinfo{pages}{041306} (\bibinfo{year}{2011})
  \dodoi{10.1103/physreve.84.041306}.

\bibitem{ohern2003jamming}
\bibinfo{author}{C.~S. O’hern}, \bibinfo{author}{L.~E. Silbert},
  \bibinfo{author}{A.~J. Liu}, and \bibinfo{author}{S.~R. Nagel},
  \enquote{\bibinfo{title}{Jamming at zero temperature and zero applied stress:
  The epitome of disorder}},  \bibinfo{journal}{Physical Review E}
  \textbf{68}(1), \bibinfo{pages}{011306} (\bibinfo{year}{2003}).

\bibitem{thompson2019critical}
\bibinfo{author}{J.~D. Thompson} and \bibinfo{author}{A.~H. Clark},
  \enquote{\bibinfo{title}{Critical scaling for yield is independent of
  distance to isostaticity}},  \bibinfo{journal}{Physical Review Research}
  \textbf{1}(1), \bibinfo{pages}{012002} (\bibinfo{year}{2019}).

\bibitem{utter2008experimental}
\bibinfo{author}{B.~Utter} and \bibinfo{author}{R.~Behringer},
  \enquote{\bibinfo{title}{Experimental measures of affine and nonaffine
  deformation in granular shear}},  \bibinfo{journal}{Physical review letters}
  \textbf{100}(20), \bibinfo{pages}{208302} (\bibinfo{year}{2008}).

\end{thebibliography}
\end{document}